\documentclass[english,aps,prd,a4paper,groupedaddress,preprintnumbers,floatfix,nofootinbib,showpacs,12pt]{revtex4}

\usepackage{graphicx}
\usepackage{subcaption}
\usepackage{enumerate}
\usepackage{amsmath}
\usepackage{amssymb}
\usepackage{amsfonts}
\usepackage{xcolor}
\usepackage{url}
\usepackage{hyperref}
\colorlet{shadecolor}{gray!15}
\definecolor{greenLinks}{rgb}{0, 0.6, 0} 
\definecolor{Links}{rgb}{0, 0, 0.6}
\definecolor{redLinks}{rgb}{0.6, 0, 0}
\definecolor{tempText}{rgb}{0.55, 0.10,0.67}
\definecolor{eprintLinks}{rgb}{0.4, 0.4, 0.4}
\definecolor{journalLinks}{rgb}{0.6, 0, 0}

\newcommand{\bi}{\begin{itemize}}
\newcommand{\ei}{\end{itemize}}

\newcommand{\beq}{\begin{equation}}
\newcommand{\eeq}{\end{equation}}
\newcommand{\bea}{\begin{eqnarray}}

\newcommand{\eea}{\end{eqnarray}}

\newcommand{\bal}{\begin{align}}
\newcommand{\eal}{\end{align}}

\newcommand{\ba}{\begin{array}}
\newcommand{\ea}{\end{array}}

\newcommand{\bset}{\begin{subequations}\begin{eqaligntwo}}
\newcommand{\eset}{\end{eqaligntwo}\end{subequations}}

\newcommand{\bseu}{\begin{subequations}\begin{eqalignno}}
\newcommand{\eseu}{\end{eqalignno}\end{subequations}}
\usepackage[font={small,it}]{caption}

\def\be{\begin{equation}}
\def\ee{\end{equation}}
\def\bea{\begin{eqnarray}}
\def\eea{\end{eqnarray}}

\newcommand{\Cinvestav}{Departamento de F\'{\i}sica, Centro de
  Investigaci{\'o}n y de Estudios Avanzados del IPN,\\ Apdo. Postal
  14-740, 07000 Ciudad de M\'exico, M\'exico.}
  
\begin{document}

\title{Effective-field theory analysis of the $\tau^- \to \eta^{(\prime)} \pi^- \nu_\tau$ decays}

\author{E. A. Garc\'es$^1$}\email{egarces@fis.cinvestav.mx} 
\author{M. Hern\'andez Villanueva$^1$}\email{emhernand@fis.cinvestav.mx} 
\author{G. L\'opez Castro$^1$}\email{glopez@fis.cinvestav.mx} 
\author{P. Roig$^1$}\email{proig@fis.cinvestav.mx} 

\affiliation{$^1$\Cinvestav}

\begin{abstract}
The rare $\tau^- \to \eta^{(\prime)}  \pi^- \nu_\tau$ decays, which are suppressed by $G$-parity in the Standard Model (SM), can be sensitive to the effects of new interactions. We 
study the sensitivity of different observables of these decays in the framework of an effective field theory that includes the most general interactions between SM fields up to 
dimension six, assuming massless neutrinos. Owing to the strong suppression of the SM isospin breaking amplitudes, we find that the different observables would
allow to set constraints on scalar interactions that are stronger than those coming from other low-energy observables.
\end{abstract}


\maketitle

\section{Introduction}
\label{sec-intro}

Rare processes are suppressed decay modes of particles originated by approximate symmetries of the SM. They provide an ideal place to look for new physics because their suppressed amplitudes 
can be of similar size as the (virtual) effects due to new particles and interactions. It turns out that having a good control of SM uncertainties is crucial to disentangle the effects of such New 
Physics contributions in precision measurements at flavor factories. 

In this paper we study the rare $\tau^- \to \eta^{(\prime)} \pi^-\nu_{\tau}$ decays, which will be forbidden if $G-$parity \cite{Lee:1956sw} were an exact symmetry of the SM ($G=Ce^{i\pi I_2}$, with 
$C$ the charge conjugation operation and $I_i$ the components of the isospin rotation operators). This process was suggested long ago \cite{Leroy:1977pq} as a clean test of Second Class Currents 
(SCC) following a classification proposed by Weinberg \cite{Weinberg:1958ut} for strangeness-conserving interactions. {According to this classification, SCC must have quantum numbers $PG(-1)^J=-1$ as opposite to (first class) currents in the SM which have  $PG(-1)^J=+1$.}
 Since isospin is only a partial symmetry of strong interactions, $G-$parity gets broken by the $u-d$ quark mass and electric 
charge differences and $\tau^- \to \eta^{(\prime)} \pi^-\nu_{\tau}$ decays can occur, although at a suppressed rate. This suppression makes interesting these decays to study the effects of 
genuine SCC, {(i. e. not induced by isospin breaking effects)}, such as the ones induced by the exchange of charged Higgs~\cite{Branco:2011iw,Jung:2010ik} or leptoquark bosons~\cite{Becirevic:2016yqi}~\footnote{Genuine SCC can also be searched for in nuclear $\beta$ decays, although having a good control of 
isospin breaking effects, which is a challenge in these processes~\cite{Severijns:2006dr} (see \cite{Triambak:2017jpw} for a recent analysis).}. We study these processes in the framework of an 
effective Lagrangian where the effects of New Physics are encoded in the most general Lagrangian involving dimension-six operators with left-handed neutrino fields.

Our study focuses on different partial and total integrated observables on $\tau^- \to \eta^{(\prime)} \pi^-\nu_{\tau}$ decays, as they can exhibit different sensitivities to the various effective 
couplings. Previous studies (including specific beyond the SM approaches) have focused mainly in the estimates of the branching fractions in the $10^{-5}\sim 10^{-6}$ ($10^{-6}\sim 10^{-8}$) 
range for the $\eta$ ($\eta'$) decay channels \cite{BRs} , as well as on the invariant mass distribution \cite{Vienna, Orsay, Escribano:2016ntp}. An important source of uncertainty in most of these 
estimates arises from the predictions used for the scalar form factor contribution. Of course, a good knowledge of the scalar form factor is necessary in order to assess the possible contributions 
of beyond SM effects. Once the $\tau^- \to \eta^{(\prime)} \pi^-\nu_{\tau}$ decays have been observed at future superflavor factories, we expect that detailed studies 
of the different observables will be very useful to disentangle the New Physics effects from the SM isospin-violating contributions~\footnote{Dedicated studies of backgrounds specific for these 
SCC decays have been carried out recently in Refs.~\cite{Guevara:2016trs, Hernandez-Tome:2017pdc}.}.

The current experimental limits for the SCC tau branching ratios of $\tau^- \to \eta  \pi^- \nu$ are: Br  $< 9.9\times 10^{-5}$, $95\%$ CL (BaBar~\cite{delAmoSanchez:2010pc}),  
 $< 7.3\times 10^{-5}$, $90\%$ CL  (Belle~\cite{Hayasaka:2009zz}) and  
 $< 1.4\times 10^{-4}$, $95\%$ CL  (CLEO \cite{Bartelt:1996iv}) collaborations, respectively.
Those upper limits lie very close to the SM estimates based on isospin breaking \cite{BRs, Vienna, Orsay, Escribano:2016ntp}. 
The corresponding BaBar limit for the $\tau^- \to \eta' \pi^- \nu_\tau$ decays is $<7.2\cdot 10^{-6}$, $95\%$ CL \cite{Aubert:2008nj}, 
while Belle obtained $<4.6\cdot 10^{-6}$, $90\%$ CL \cite{Hayasaka:2009zz} (CLEO set the earlier upper bound $7.4\times10^{-5}$, $90\%$CL \cite{Bergfeld:1997zt}). 
Future experiments at the intensity frontier like Belle II \cite{Abe:2010gxa}, which will accumulate $4.5\times 10^{10}$ tau lepton pairs in the full dataset, are expected to provide the first 
measurements of the $\tau^- \to \eta^{(\prime)} \pi^-\nu_{\tau}$ SCC decays \cite{B2TIPReport}.
 
This paper is organized as follows: in section \ref{eff-theory} we set our conventions for the effective field theory analysis of the $\tau^- \to \nu_{\tau} \bar{u}d$ decays, to be used in the remainder 
of the article. In section \ref{Amplitudes}, we discuss the different effective weak currents contributing to the considered decays and define their corresponding hadronic form factors. The tensor form 
factor within low-energy QCD is computed in section \ref{tensor}. In section \ref{Observables} we discuss the different observables that can help elucidating non-SM contributions to the 
$\tau^- \to \eta^{(\prime)} \pi^-\nu_{\tau}$ decays and in section \ref{Concl} we state our conclusions.
  
\section{Effective theory analysis of $\tau{^-} \to \nu_{\tau} \bar{u}d$} \label{eff-theory}

The effective Lagrangian with $SU(2)_L\otimes U(1)$ invariant dimension six operators at the weak scale contributing to low-energy charged current processes\footnote{The most general effective 
Lagrangian including SM fields was derived in Refs.~\cite{Buchmuller:1985jz, Grzadkowski:2010es}.} can be written as~\cite{Bhattacharya:2011qm, Cirigliano:2009wk}
\bea
{\cal L}^{(\rm{eff})}
= {\cal L}_{\rm{SM}} + \frac{1}{\Lambda^2} \sum_{i} ~ \alpha_i O_i \ \longrightarrow \ 
{\cal L}_{\rm{SM}} +  \frac{1}{v^2}  \, \sum_{i}  \, \hat{\alpha}_i   ~ O_i \, ,
\eea
 with $ \hat{\alpha}_i = (v^2/\Lambda^2) \alpha_i$ the dimensionless new physics couplings, which are $O(10^{-3})$  for an scale  $\Lambda \sim 1\ {\rm TeV}$.

The low-scale  $O(1 \ {\rm GeV})$  effective Lagrangian for semi-leptonic ($\ell\,=\,e,\,\mu,\,\tau$) strangeness {and lepton-flavor} conserving transitions \footnote{Strangeness-changing processes are 
discussed in an EFT framework in Refs.~\cite{Chang:2014iba, Gonzalez-Alonso:2016etj, Gonzalez-Alonso:2016sip}.} involving only left-handed neutrino fields is given by 
(subscripts $L(R)$ refer to left-handed (right-handed) chiral projections)

{
\bea
{\cal L}_{\rm CC}
&=&  \frac{-4G_F}{\sqrt{2}} \, V_{ud} \, \Bigg[
 \Big(1 + [v_L]_{\ell \ell } \Big) \   \bar{\ell}_L \gamma_\mu  \nu_{\ell L}    \ \bar{u}_L \gamma^\mu d_L
 \ + \  [v_R]_{\ell \ell }  \   \bar{\ell}_L \gamma_\mu  \nu_{\ell L}    \ \bar{u}_R\gamma^\mu d_R
\nonumber\\
&+&  [s_L]_{\ell \ell }  \   \bar{\ell}_R   \nu_{\ell L}    \ \bar{u}_R  d_L
\ + \  [s_R]_{\ell \ell }  \   \bar{\ell}_R   \nu_{\ell L}    \ \bar{u}_L d_R
\nonumber \\
& + &   [t_L]_{\ell \ell }  \   \bar{\ell}_R   \sigma_{\mu \nu} \nu_{\ell L}    \ \bar{u}_R  \sigma^{\mu \nu} d_L
\Bigg]~+~{\rm h.c.}~,
\label{eq:leffq}
\eea}
where  
$G_F$ stands for the tree-level definition of the Fermi constant, $\sigma^{\mu \nu} \equiv i [\gamma^\mu, \gamma^\nu]/2$, and $v_L = v_R = s_L = s_R = t_L = 0$ gives the SM Lagrangian.
{In the Lagrangian above, as usual, Higgs, $W^{\pm}$, and $Z$ boson degrees of freedom have been integrated out, as well as $c$, $b$ and $t$ quarks.} Since 
we will be considering only CP-even observables, the effective couplings $v_{L,R}$,  $s_{L,R}$, and $t_L$ characterizing New Physics\footnote{These couplings, as functions of the $\hat{\alpha}_i$ couplings of the SM electroweak 
gauge invariant weak-scale operators, can be found in appendix A of Ref.~\cite{Bhattacharya:2011qm}.} can be taken real.

 In terms of equivalent effective couplings\footnote{The physical amplitudes are renormalization scale and scheme independent. However, the individual 
effective couplings $\epsilon_{i}$ and hadronic matrix elements do depend on the scale. As it is conventionally done, we choose $\mu=2$ GeV in the $\overline{MS}$ scheme.} ($\epsilon_{L,R}=v_{L,R},\ \epsilon_S=s_L+s_R, \ \epsilon_P=s_L-s_R$ and $\epsilon_T=t_L$) we have the following form of the 
semileptonic effective Lagrangian\footnote{The factor 2 in the tensor contribution originates from the identity $\sigma^{\mu\nu}\gamma^5 = - \frac{i}{2} \epsilon^{\mu\nu\alpha\beta}\sigma_{\alpha\beta}$.} 
{(particularized for $\ell=\tau$)}:
\bea
{\cal L}_{\rm CC} 
&=&
- \frac{G_F V_{ud}}{\sqrt{2}} \,
\Big(1 + \epsilon_L + \epsilon_R  \Big)
\Bigg[
\bar{\tau}  \gamma_\mu  (1 - \gamma_5)   \nu_{\tau} 
\cdot \bar{u}   \Big[ \gamma^\mu \ - \  \big(1 -2  \widehat{\epsilon}_R  \big)  \gamma^\mu \gamma_5 \Big] d \nonumber\\
&+& \bar{\tau}  (1 - \gamma_5) \nu_{\tau}
\cdot \bar{u}  \Big[  \widehat{\epsilon}_S  -  \widehat{ \epsilon}_P \gamma_5 \Big]  d
+2 \widehat{\epsilon}_T     \,   \bar{\tau}   \sigma_{\mu \nu} (1 - \gamma_5) \nu_{\tau}    \cdot  \bar{u}   \sigma^{\mu \nu} d
\Bigg]+{\rm h.c.}, \ \ \ \  \
\label{eq:leffq2} 
\eea
where $\widehat{\epsilon_i}\equiv \epsilon_i/(1+\epsilon_L+\epsilon_R)$ for $i=R, S, P, T$. This factorized form is useful as long as conveniently normalized rates allow to cancel the overall factor $(1+\epsilon_L+\epsilon_R)$. 
Keeping terms linear in the small effective couplings, the $\widehat{\epsilon}_i$'s reduce to the expression in Ref. \cite{Bhattacharya:2011qm}. 

\section{Semileptonic $\tau$ decay amplitude}
\label{Amplitudes}

Let us consider the semileptonic $\tau^-(p) \to \eta^{(\prime)}(p_{\eta}) \pi^-(p_{\pi})\nu_{\tau}(p')$ decays. Owing to the parity of pseudoscalar mesons, only the vector, scalar and tensor currents give a non-zero 
contribution to the decay amplitude, which reads~\footnote{The short-distance electroweak radiative corrections encoded in $S_{EW}$~\cite{Erler:2002mv} do not affect the scalar and tensor contributions. However, the error made by 
taking $\sqrt{S_{EW}}$ as an overall factor in eq.~(\ref{eq:amp}) is negligible.}:
\begin{eqnarray} \label{eq:amp}
{\cal M} &=&  {\cal M}_{V} +{\cal M}_{S} +{\cal M}_{T} \nonumber \\
& =& \frac{G_F V_{ud} \sqrt{S_{EW}}}{\sqrt{2}}(1 + \epsilon_L +  \epsilon_R ) \, \left[ 
 L_\mu  H^{\mu}
+   \widehat{\epsilon}_{S}  L  H + 2 \widehat{\epsilon}_T  L_{\mu\nu}\, H^{\mu\nu}  \right ], 
\end{eqnarray}
where we have defined the following leptonic currents 
\begin{eqnarray}
L_{\mu} &=& \bar{u} (p')  \gamma_\mu  (1-\gamma_5) u(p),  \nonumber \\
 L&=&\bar{u} (p')  (1+\gamma_5)  u(p), \\ 
L_{\mu\nu}&=&\bar{u} (p')  \sigma_{\mu \nu} (1 + \gamma_5)  u(p), \nonumber 
\end{eqnarray}

In eq. (\ref{eq:amp}) we have defined the following vector, scalar and tensor hadronic matrix elements 
\begin{eqnarray}\label{eq:hadME}
H^\mu &=& {\langle \eta^{(\prime)} \pi^-}|  \bar{d}   \gamma^\mu  u {| 0 \rangle}=  c_V  Q^{\mu} F_+(s) +  c_S \frac{ \Delta^{QCD}_{K^0K^+}}{s} q^{\mu} F_0(s),  \label{had-v}\\
H &=& {\langle \eta^{(\prime)} \pi^- | } \bar{d} u{| 0 \rangle}=  F_S(s), \label{had-s} \\ 
H^{\mu\nu} &=& {\langle \eta^{(\prime)} \pi^- | } \bar{d}   \sigma^{\mu \nu} u {| 0 \rangle}  =  i F_T(s) (p_{\eta}^\mu p_{\pi}^\nu - p_{\pi}^\mu p_{\eta}^\nu)  \label{had-t},
\end{eqnarray}
where we have defined  $q^{\mu}= (p_{\eta^{(\prime)}} + p_{\pi})^\mu$, $Q^{\mu}= (p_{\eta^{(\prime)}}-p_{\pi})^\mu + (\Delta_{\pi^-\eta^{(\prime)}}/s) q^{\mu}$,  $s=q^2$ and $\Delta_{ij}\equiv m_i^2-m_j^2$, 
$\Delta^{QCD}_{K^0K^+}={m_{K^0}}^2 - m_{K^{+}}^2 + m_{\pi^{+}}^2 - m_{\pi{^0}}^2$; the constants $c_S=\sqrt{\frac{2}{3}},~c_V=\sqrt{2}$, denote Clebsch-Gordan flavor coefficients. In the $\eta'$ case 
$c_S=\frac{2}{\sqrt{3}}$ ($c_V$ remains to be $\sqrt{2}$). For simplicity we have not written the labels in the  $F_{+,0,S,T}$ form factors, which are different for specific hadronic channels. 

The divergence of the vector current relates the $F_S(s)$ and $F_0(s)$ form factors via
\begin{equation} \label{sff-relation}
F_S(s)=c_S \frac{\Delta^{QCD}_{K^0K^+}}{(m_d-m_u)}F_0(s)\,.
\end{equation}
Since \cite{ChPT}
\begin{equation}
 \frac{\Delta^{QCD}_{K^0K^+}}{(m_d-m_u)}\,=\,B\left(1-\frac{1}{4}\frac{m_u-m_d}{m_s-\hat{m}}\right)\sim B\,,
\end{equation}
where $\hat{m}\equiv(m_u+m_d)/2$ and $BF^2=<0|\bar{q}q|0>\sim -(270$ MeV$)^3$ \cite{Aoki:2016frl}, it is seen --by using $F\sim92$ MeV-- that $B\sim M_\tau$. Thus, $F_S(s)$ basically inherits 
the strong isospin suppression of $F_0(s)$.

Observe that the scalar contribution in eq. (\ref{had-s}) can be `absorbed' into the vector current amplitude by using the Dirac equation  $ L=L_\mu q^\mu/M_\tau $ and eq. (\ref{sff-relation}).  
This can be achieved by replacing 
\begin{equation}\label{s-dependence epsilon_S}
 c^S \frac{\Delta_{K^0K^+}^{QCD}}{s}\ \longrightarrow\  c^S \frac{\Delta_{K^0K^+}^{QCD}}{s} \left[1+\frac{s\widehat{\epsilon}_S}{m_{\tau}(m_d-m_u)} \right]\ ,
\end{equation}
in the second term of eq. (\ref{had-v}).
 We will see in the next section that the remaining contribution to eq.~(\ref{eq:amp}), 
given by the tensor current (${\cal M}_{T}$), is also suppressed in low-energy QCD.

\section{Hadronization of the tensor current} \label{tensor}
The hadronization of the tensor current, eq. (\ref{had-t}), is one of the most difficult inputs to be reliably estimated. In the tau lepton decays under consideration, the momentum transfer ranges within $(m_{\eta^{(\prime)}}+m_{\pi})^2 \leq s \leq M_\tau^2$, 
which is the kinematic region populated by light resonances. Here we will neglect {the} $s$-dependence, namely $F_T^{\pi\eta^{(\prime)}}(s)=
F_T^{\pi\eta^{(\prime)}}(0)\equiv F_T^{\pi\eta^{(\prime)}} $, and we will estimate its value using Chiral 
Perturbation Theory \cite{Weinberg:1978kz, Gasser:1983yg, Gasser:1984gg, Colangelo:1999kr}. We do not consider tensor current contributions at the 
next-to-leading chiral order in order to keep predictability.

{A comment is in order with respect to neglecting resonance contributions in the hadronization of the tensor current, as it couples to the $J^{PC}=1^{--}$ 
resonances, being the $\rho(770)$ its lightest representative. In principle, one should expect a contribution from these resonances to the considered decays, 
providing an energy-dependence to $F_T$ and increasing its effect in the observables that we study. The $\rho(770)$ will contribute very little to the $\eta'\pi$ decay 
mode, owing to kinematical constraints, and the contributions of $\rho(1450)$ and $\rho(1700)$ will be damped by phase space and their wide widths. Thus, it is quite justified to 
assume $F_T^{\pi\eta^{\prime}}(s)= F_T^{\pi\eta^{\prime}}(0)\equiv F_T^{\pi\eta^{\prime}} $. Our previous reasoning does not apply to the vector resonance 
contribution to $F_T^{\pi\eta}(s)$, however. It is predicted by large-$N_C$ arguments that vector resonances couple to the tensor current with 
a strength only a factor $1/\sqrt{2}$ smaller than to the vector current \cite{Cata:2008zc} (which is also supported by lattice evaluations 
\cite{Becirevic:2003pn, Braun:2003jg, Donnellan:2007xr}). Consequently, the $\rho(770)$ contribution to $F_T^{\pi\eta}(s)$ should not be negligible 
(the vector current contribution of the $\rho(770)$ state to the $\tau^-\to\eta\pi^-\nu_\tau$ branching ratio is $\sim 1/6$, according to 
Ref.~\cite{Escribano:2016ntp}). As a result, our limits on the allowed values of $\widehat{\epsilon_T}$ obtained from the $\pi\eta$ decay mode, which are 
presented in the next section, could be made stronger including this missing contribution. However, as we will see, the main point of this article is that 
$\tau^-\to\eta^{(\prime)}\pi^-\nu_\tau$ decays are competitive setting limits on non-standard scalar interactions in charged current decays, while they are 
not in tensor interactions \footnote{{As we discuss at the end of section \ref{Concl}, our upper limit on $\widehat{\epsilon_T}$ is $\sim 0.5$, 
while the $10^{-4}$ level is reached in radiative pion decays. Our educated guess for the $\rho(770)$ contribution through the tensor current to the 
$\tau^-\to\eta\pi^-\nu_\tau$ decays (based on its contribution through the vector current) is that with a good understanding of the former we could probably 
reach $\widehat{\epsilon_T}\lesssim 10^{-2}$, but not the $10^{-4}$ level.}}. 
This main conclusion is not affected by our assumption $F_T^{\pi\eta}(s)=F_T^{\pi\eta}(0)\equiv F_T^{\pi\eta}$.}
Therefore our analyses (right panel in figures \ref{fig:Del} and \ref{fig:Delp}) involving the tensor source with a constant form factor should be simply viewed as a benchmark to compare with those with the scalar source, and not as a full fledged and theoretically sound computation.

According to Ref.~\cite{Cata:2007ns}, there are only four operators at the leading chiral order, $\mathcal{O}(p^4)$, that include the tensor current. Only the operator with coefficient $\Lambda_2$ contributes to the decays we 
are considering~\footnote{We note that although $SU(3)$ flavor symmetry was considered in Ref.~\cite{Cata:2007ns}, extending it 
to U(3) (for a consistent treatment of the $\eta^\prime$ meson) does not bring any extra operator at this order, as this extension entails the appearance 
of a $\log(\det[u])$ factor, which adds $\mathcal{O}(p^2)$ to the chiral counting, belonging thus to the next-to-leading order Lagrangian that we do not 
consider. Also, odd-intrinsic parity sector operators including the tensor source first appear at $\mathcal{O}(p^8)$ \cite{Cata:2007ns}.}:
\begin{equation}\label{eq:LagrangianTSource}
 \mathcal{L} = \Lambda_1 \left\langle t_+^{\mu\nu} f_{+\,\mu\nu}\right\rangle-i \Lambda_2 \left\langle t_+^{\mu\nu} u_\mu u_\nu \right\rangle+...\,
\end{equation}
where $t_+^{\mu\nu}=u^\dagger t^{\mu\nu} u^\dagger + u {t^{\mu\nu}}^\dagger u$ and $\left\langle ... \right\rangle$ stands for a trace in flavor space. 
The chiral tensors entering eq.~(\ref{eq:LagrangianTSource}) are $u_\mu = i \left[ u^\dagger (\partial_\mu - i r_\mu)u - u (\partial_\mu-i \ell_\mu) u^\dagger\right]$, 
including the left- and right-handed sources $\ell_\mu$ and $r_\mu$, the (chiral) tensor sources, $t^{\mu\nu}$ and its adjoint; and $f_+^{\mu\nu}= u F_L^{\mu\nu} u^\dagger + u^\dagger F_R^{\mu\nu} u$, including the 
field-strength tensors for $\ell_\mu$ and $r_\mu$.

The non-linear representation of the pseudoGoldstone bosons is given by $u=\mathrm{exp}\left\lbrace \frac{i}{\sqrt{2}F}\phi\right\rbrace$, where 
\[
 \phi = \begin{pmatrix} \frac{\pi^3+ \eta_q}{\sqrt{2}} & \pi^+ & K^+ \\ \pi^- & \frac{-\pi^3+ \eta_q}{\sqrt{2}} & K^0\\ K^- & \overline{K^0} & \eta_s  \end{pmatrix} \,,
\]
with $\eta_q = C_q \eta + C_{q^\prime} \eta^\prime$ and $\eta_s = -C_s \eta + C_{s^\prime} \eta^{\prime}$ the light and strange quark components of the $\eta,\ \eta'$ mesons, respectively ($\pi^3$ 
is the pseudoGoldstone having the flavor quantum numbers of the $\lambda_3$ Gell-Mann matrix, which coincides with the $\pi^0$ neglecting isospin breaking). These constants describing the 
$\eta-\eta^\prime$ mixing are given by \cite{ChPTLargeN}
\small{
\begin{eqnarray}\label{definitions_Cq_Cs}
 C_q & \equiv & \frac{F}{\sqrt{3}\mathrm{cos}(\theta_8-\theta_0)}\left(\frac{\mathrm{cos}\theta_0}{f_8}-\frac{\sqrt{2}\mathrm{sin}\theta_8}{f_0}\right)\,,\quad C_{q^\prime}\equiv\frac{F}{\sqrt{3}\mathrm{cos}(\theta_8-\theta_0)}\left(\frac{\sqrt{2}\mathrm{cos}\theta_8}{f_0}+\frac{\mathrm{sin}\theta_0}{f_8}\right)\,,\nonumber\\
 C_s & \equiv & \frac{F}{\sqrt{3}\mathrm{cos}(\theta_8-\theta_0)}\left(\frac{\sqrt{2}\mathrm{cos}\theta_0}{f_8}+\frac{\mathrm{sin}\theta_8}{f_0}\right)\,,\quad C_{s^\prime}\equiv\frac{F}{\sqrt{3}\mathrm{cos}(\theta_8-\theta_0)}\left(\frac{\mathrm{cos}\theta_8}{f_0}-\frac{\sqrt{2}\mathrm{sin}\theta_0}{f_8}\right)\, ,
\end{eqnarray}
}
and the corresponding values of the pairs of decay constants and mixing angles are \cite{2anglemixing}
\begin{equation}\label{values_mixingangless&decayscts}
 \theta_8\,=\,\left(-21.2\pm1.6\right)^\circ,\quad  \theta_0\,=\,\left(-9.2\pm1.7\right)^\circ,\quad f_8\,=\, \left(1.26\pm0.04\right)F,\quad f_0\,=\, \left(1.17\pm0.03\right)F\,
\end{equation}
with $F\sim 92.2$ MeV being the pion decay constant.

We recall \cite{Cata:2007ns} that the tensor source ($\bar{t}^{\mu\nu}$) is related to its chiral projections ($t^{\mu\nu}$ and ${t^{\mu\nu}}^\dagger$) by 
means of
\begin{eqnarray}
t^{\mu\nu}\,=\,P_L^{\mu\nu\lambda\rho} \bar{t}_{\lambda\rho}\,,\;\; 4 P_L^{\mu\nu\lambda\rho}\,=\, (g^{\mu\lambda}g^{\nu\rho}-g^{\mu\rho}g^{\nu\lambda}+i \epsilon^{\mu\nu\lambda\rho})\,,
\end{eqnarray}
with $\bar{\Psi} \sigma_{\mu\nu} \bar{t}^{\mu\nu} \Psi$ as the tensor current.

Taking the functional derivative of eq.~(\ref{eq:LagrangianTSource}) with respect to $\bar{t}_{\alpha\beta}$, putting all other external sources to zero, 
expanding $u$ and taking the suitable matrix element, it can be shown that in the limit of isospin symmetry 

\begin{equation}\label{isym-couplings}
 i \left\langle \pi^-\pi^0\Bigg|\frac{\delta\mathcal{L}_{\chi PT}^{\mathcal{O}(p^4)}}{\delta \bar{t}_{\alpha\beta}}\Bigg|0\right\rangle\,=\,\frac{\sqrt{2}\Lambda_2}{F^2}(p_-^\alpha p_0^\beta-p_0^\alpha p_-^\beta)
 \,,\;\; \left\langle \pi^-\eta^{(\prime)}\Bigg|\frac{\delta\mathcal{L}_{\chi PT}^{\mathcal{O}(p^4)}}{\delta \bar{t}_{\alpha\beta}}\Bigg|0\right\rangle\,=\,0\,.
\end{equation}
Once isospin symmetry breaking is taken into account, the leading contributions to the tensor hadronic matrix elements are given by:
\begin{eqnarray}\label{ib-couplings}
 i \left\langle \pi^-\pi^0\Bigg|\frac{\delta\mathcal{L}_{\chi PT}^{\mathcal{O}(p^4)}}{\delta \bar{t}_{\alpha\beta}}\Bigg|0\right\rangle\,&=&\,\frac{\sqrt{2}\Lambda_2}{F^2}(p_-^\alpha p_0^\beta-p_0^\alpha p_-^\beta)\ , \\
 \;\; i \left\langle \pi^-\eta^{(')}\Bigg|\frac{\delta\mathcal{L}_{\chi PT}^{\mathcal{O}(p^4)}}{\delta \bar{t}_{\alpha\beta}}\Bigg|0\right\rangle\,&=& \epsilon_{\pi \eta^{(')}} \frac{\sqrt{2}\Lambda_2}{F^2}(p_\pi^\alpha p_\eta^\beta-p_\eta^\alpha p_\pi^\beta)\,. \label{ib-couplings2}
\end{eqnarray} 
For the numerical values of the isospin breaking mixing parameters we will take the determinations $\epsilon_{\pi\eta}=(9.8\pm 0.3)\cdot 10^{-3}$ and 
$\epsilon_{\pi\eta^\prime}=(2.5\pm 1.5)\cdot 10^{-4}$ ~\cite{Escribano:2016ntp}. To our knowledge, there is no phenomenological or theoretical information on 
$\Lambda_2$. However, $\Lambda_1$ appearing in the Lagrangian eq.~(\ref{eq:LagrangianTSource}) was predicted --using QCD short-distance constraints-- in 
Ref.~\cite{Mateu:2007tr} to be 
\begin{equation}
  \Lambda_1 = \frac{<0|\bar{q}q|0>}{M_V^2}\sim (33\pm2)\,\mathrm{MeV}\,,
\end{equation}
where we took $<0|\bar{q}q|0>$ from \cite{Aoki:2016frl}. This yields $\frac{\Lambda_1}{4\pi F}=0.028\pm0.002$, which is consistent with the chiral counting proposed in 
Ref.~\cite{Cata:2007ns}. As a conservative estimate~\footnote{We note that the operators with coefficients $\Lambda_1$ and $\Lambda_2$ in eq. 
(\ref{eq:LagrangianTSource}) share the same chiral counting order~\cite{Cata:2007ns}.}, we will assume $\frac{|\Lambda_2|}{4\pi F}\leq 0.05$ in our analysis.  
This, in turn, results in $|F_T^{\pi\eta}|\leq 0.094 ~\text{GeV}^{-1}$ and $|F_T^{\pi\eta^\prime}|\leq 2.4\cdot10^{-3} 
~\text{GeV}^{-1}$ (we note that, according to our definition in eq.~(\ref{had-t}), $F_T^{\pi\eta^{(\prime)}}$ includes the factor $\epsilon_{\pi\eta^{(\prime)}}$. If, instead, the tilded 
form factors of Ref.~\cite{Escribano:2016ntp} are used, then $\Big|\widetilde{F_T}^{\pi\eta}\Big|=\Big|\widetilde{F_T}^{\pi\eta^{(\prime)}}\Big|=\frac{\sqrt{2}\Lambda_2}{F^2}\lesssim 9.59$ GeV$^{-1}$). 
{Our uncertainty in the sign of $F_T$ translates in the corresponding lack of knowledge for the interference between tensor and scalar or vector contributions.}
{We finally note that the overall suppression given by the $\epsilon_{\pi \eta^{(')}}$ factors in eq.~(\ref{ib-couplings2}), together with our 
estimate of $|\Lambda_2|$, make $\tau^- \to \pi^-\eta^{(\prime)} \nu_{\tau}$ decays not competitive with the radiative pion decay in setting bounds on non-standard 
tensor interactions.}

\section{Decay observables}\label{Observables}

 Most of the existing studies of $\tau^- \to \pi^-\eta^{(\prime)} \nu_{\tau}$ decays have focused on the branching ratio \cite{BRs} and only a few of them have provided predictions for the spectra 
 in the invariant mass of the hadronic system \cite{Vienna, Orsay, Escribano:2016ntp}. Once these $G-$parity forbidden decays have been discovered at Belle II, the next step will be to 
characterize their hadronic dynamics and to look for possible effects of \textit{genuine} SCC (New Physics). This will require the use of more detailed observables like the hadronic 
spectrum and angular distributions or Dalitz plot analyses. In this section we focus in the decay observables that can be accessible in the presence of New Physics characterized by the effective 
weak couplings described in Section \ref{eff-theory}.

In the rest frame of the $\tau$ lepton, the differential width for the $\tau^- \to \pi^-\eta^{(\prime)} \nu_{\tau}$ decay is 
\begin{eqnarray}\label{dalitz}
\frac{d^2\Gamma}{dsdt} = \frac{1}{32 (2\pi)^3M_\tau^3} \overline{|{\cal M}|^2}\ ,
\end{eqnarray}
where $\overline{|{\cal M}|^2}$ is the unpolarized spin-averaged squared matrix element, $s$ is the invariant mass of the $\eta^{(\prime)}\pi^-$ system (taking values within $(m_{\eta^{(\prime)}}+m_{\pi})^2 \leq s 
\leq M_\tau^2$) and $t= (p'+p_{\eta^{(\prime)}})^2=(p-p_{\pi^-})^2$ with kinematic limits given by $t^{-}(s) \leq t \leq t^+(s)$, and
\begin{eqnarray}
t^{\pm}(s)= \frac{1}{2s} \left[  2s(M_\tau^2+m_{\eta^{(\prime)}}^2-s){ - (M_\tau^2 - s)(s + m_{\pi}^2 - m_{\eta^{(\prime)}}^2)} \pm (M_\tau^2-s)\sqrt{ \lambda(s, m_{\pi} ^2,m_{\eta^{(\prime)}}^2 )}\right]\ ,
\end{eqnarray}
where the Kallen function is defined as  $\lambda(x,y,z) =  x^2+y^2+z^2 -2xy -2xz - 2yz$. 
\subsection{Dalitz plot}
The unpolarized spin-averaged squared amplitude in the presence of New Physics interactions is given by
\begin{eqnarray}\label{M2}
 \overline{|{\cal M}|^2}= \frac{2  G_F^2 |V_{ud}|^2 S_{EW}}{s^2}(1+\epsilon_L+\epsilon_R)^2 \left(   M_{0+} +  M_{T+}  + M_{T0}  + M_{00} + M_{++} + M_{TT} \right)
 \end{eqnarray}
 where $M_{00}$ , $M_{++}$ and $M_{TT}$ originate from the scalar, vector and tensor contributions to the amplitude respectively, 
 and $M_{0+}$,  $M_{T+}$,  $M_{T0}$ are their corresponding interference terms. Their expressions are 
\small{
\begin{eqnarray}\label{M2contributions}
M_{0+} &=& 2  c_Vc_S m_\tau^2  \times {\rm Re}[F_+(s)F^*_0(s)] \Delta^{QCD}_{K^0K^+}
 \left( 1 + \frac{  \widehat{\epsilon}_S s}{ m_\tau(m_d- m_u)} \right) \nonumber \\
&&\ \ \ \ \  \times\left(   s(m_{\tau}^2-s+\Sigma_{\pi\eta^{(')}}-2t) + m_{\tau}^2\Delta_{\pi\eta^{(')}}  \right)   \,,\nonumber \\
M_{T+} &=&  - 4  c_V  \widehat{\epsilon}_T  m_\tau^3 s {\rm Re}[F_{T}F_{+}^*(s)]    
\left(1-\frac{s}{ m_\tau^2} \right)\lambda(s,m_{\pi}^2, m_{\eta^{(')}}^2) \,, \nonumber \\
M_{T0}&=& - 4 c_S \Delta^{QCD}_{K^0K^+}\widehat{\epsilon}_T  m_\tau s  {\rm Re}[F_T F_{0}^*(s)]  \left( 1 + \frac{  \widehat{\epsilon}_S s}{ m_\tau(m_d- m_u)} \right)  \nonumber \\ 
&& \ \ \ \ \ \times \left(s(m_{\tau}^2-s-2t+\Sigma_{\pi\eta^{(')}})+m_{\tau}^2 \Delta_{\pi\eta^{(')}} \right)  \,, \nonumber \\ 
M_{00}&=&  c_S^2{(\Delta^{QCD}_{K^0K^+})^2}m_\tau^4 \left(1-\frac{s}{m_\tau^2} \right)  \left|F_0(s)\right|^2 \left( 1+\frac{  \widehat{\epsilon}_S s}{  m_\tau( m_d- m_u)}\right)^2 \,,\nonumber  \\ 
M_{++} &=&c_V^2  |F_{+}(s)|^2  \left[ m_\tau^4 (s+\Delta_{\pi\eta^{(')}})^2 -m_\tau^2 s \left( 2 \Delta_{\pi\eta^{(')}}(s+2t-2m_\pi^2)+\Delta_{\pi\eta^{(')}}^2+s(s+4t)\right) \right. \nonumber \\ 
 &&\ \ \ \ \ + \left. 4 m_{\eta^{(')}}^2 s^2 (m_\pi^2-t) +4 s^2 t (s+t-m_\pi^2)\right]  \,, \nonumber \\
M_{TT} &=& 4  \widehat{\epsilon}_T^2 F_{T}^2 s^2 \left[ \frac{}{} m_{\eta^{(')}}^4 (m_\tau^2-s)-2 m_{\eta^{(')}}^2 (m_\tau^2-s)(s+2t-m_\pi^2)-m_\pi^4 (3 m_\tau^2+s) \right.  \nonumber \\ 
&&\ \ \ \ \  \left. + 2 m_\pi^2\left((s+m_\tau^2)(s+2t)-2m_\tau^4\right)-s\left((s+2t)^2-m_\tau^2(s+4t)\right)
 \frac{}{}  \right]  \ , 
\end{eqnarray}
}
where we have defined $\Delta_{\pi\eta^{(')}}=m_{\pi^-}^2-m_{\eta^{(')}}^2$, $\Sigma_{\pi\eta^{(')}}=m_{\pi^-}^2+m_{\eta^{(')}}^2$. 

\begin{figure}[h!]
\begin{center}
\includegraphics[width=.496\textwidth]{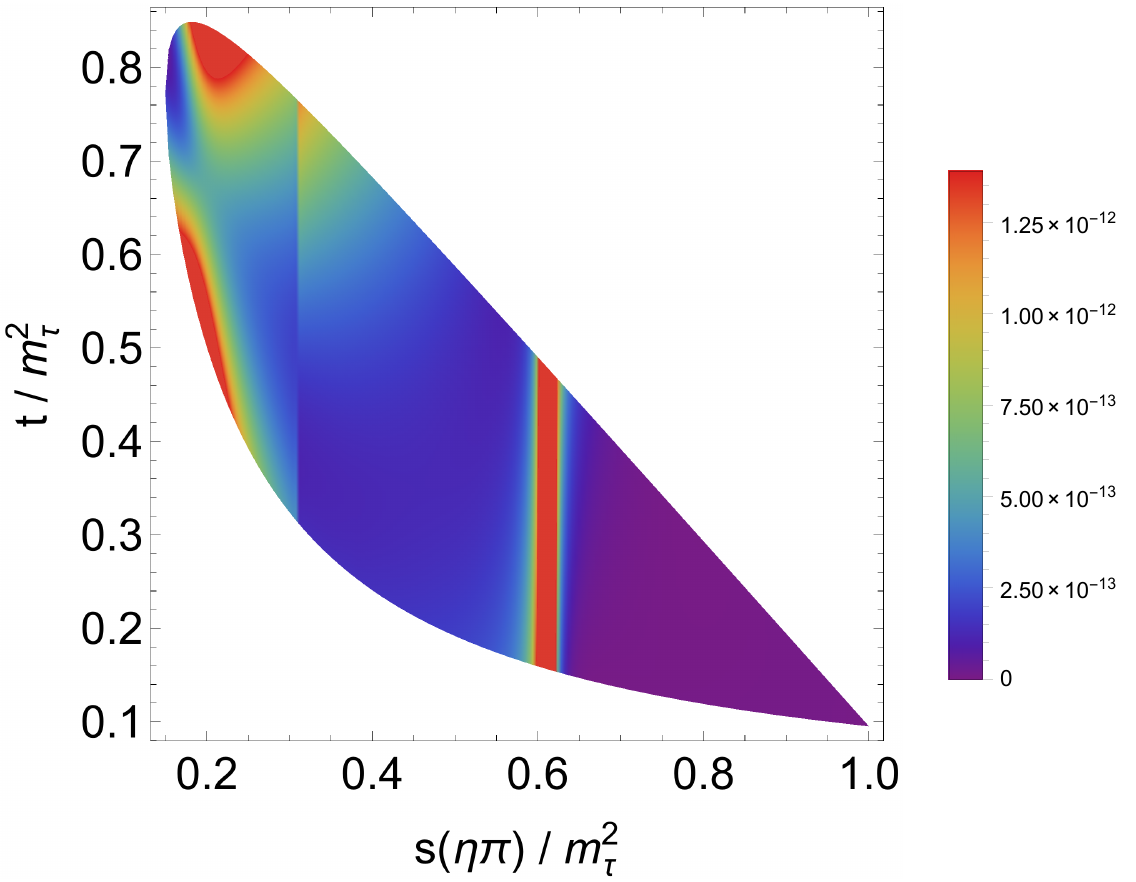}
\includegraphics[width=.496\textwidth]{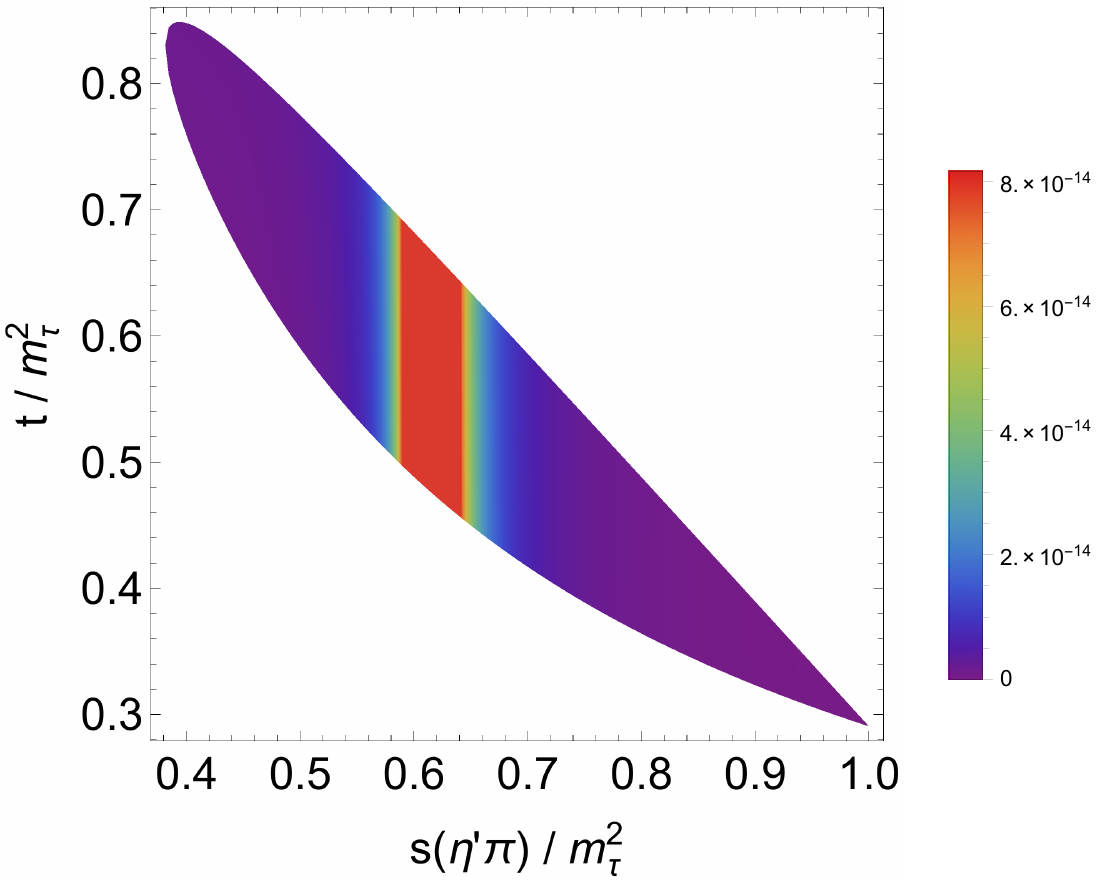}
\includegraphics[width=.496\textwidth]{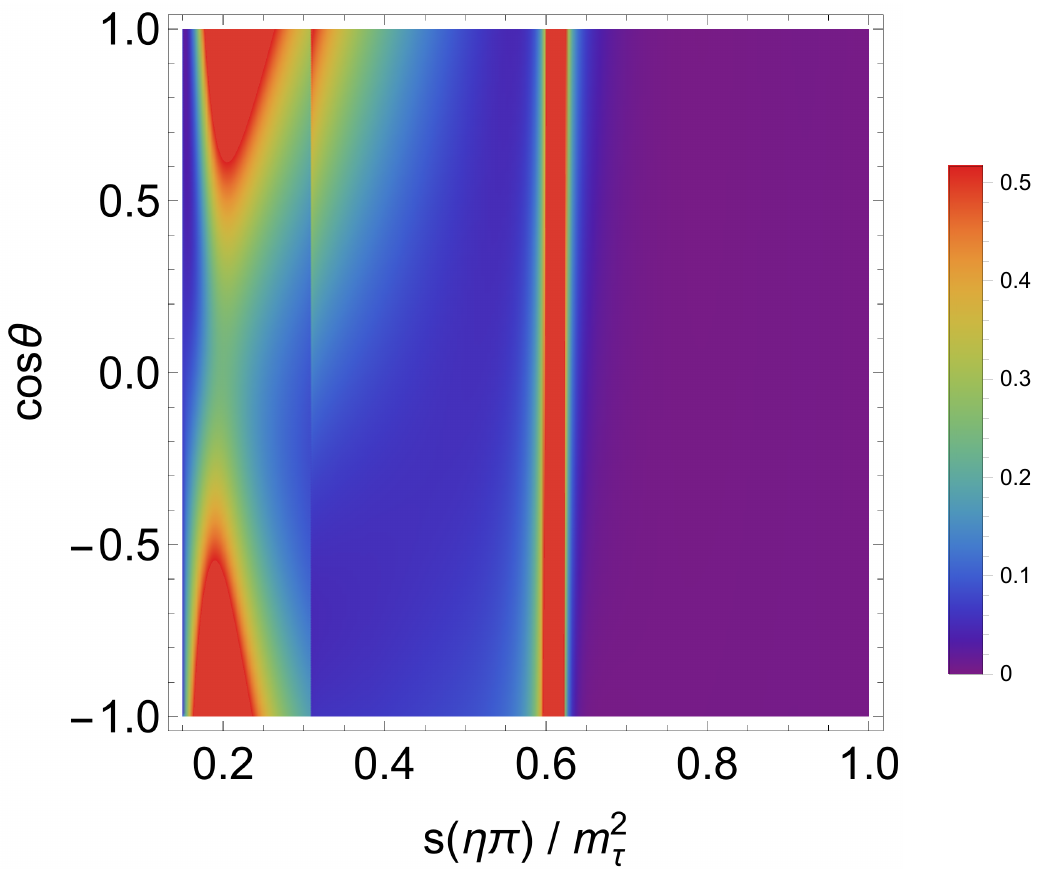}
\includegraphics[width=.496\textwidth]{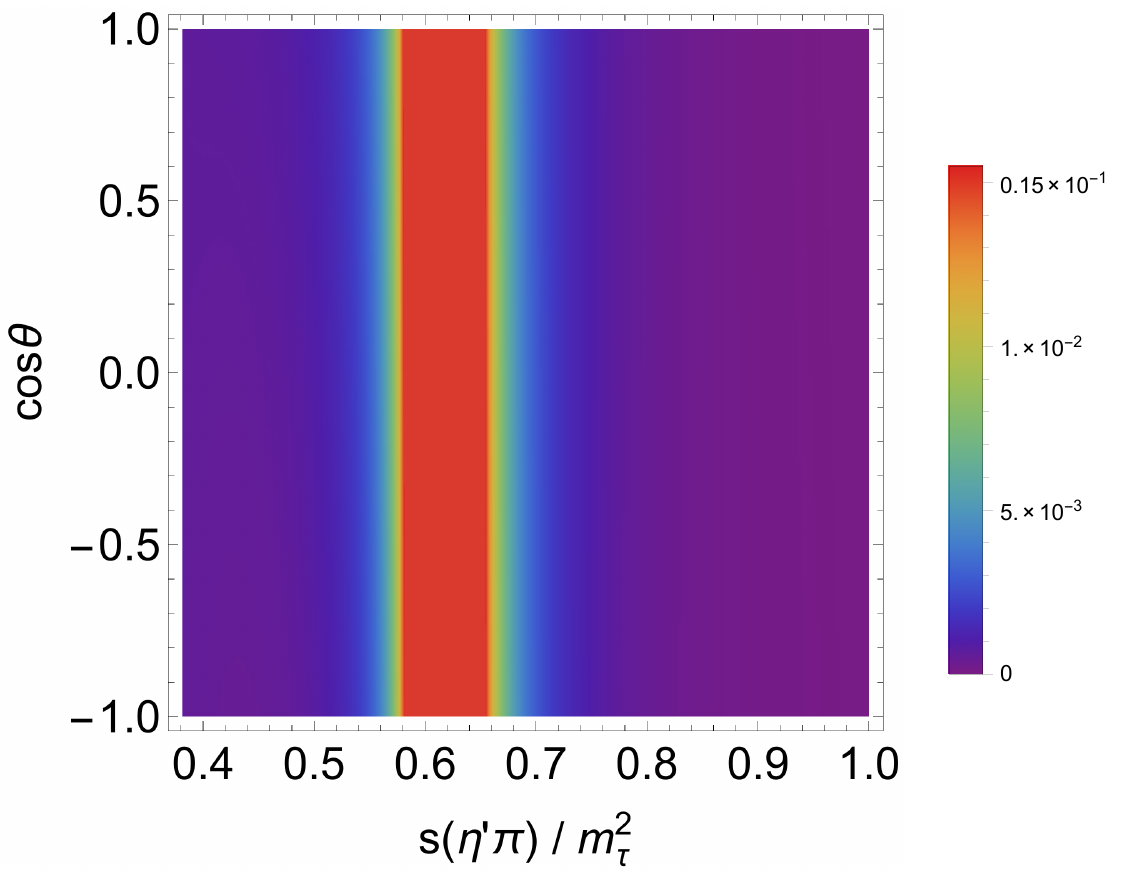}
\end{center}
  \caption{\label{fig:DalSM} Dalitz plot distribution $\overline{|{\cal M}|^2}_{00}$ in the SM, eq. (\ref{M2}): the $\eta\pi$ ($\eta'\pi$) case is shown in the left (right) 
  column. The figures in the second row show the double differential decay distribution in the $(s, \cos \theta)$ variables according to eq.~(\ref{angulardistribution}) 
  {normalized to the tau width}, for both decay channels. 
  {The Mandelstam variables, $s$ and $t$, are normalized to $M_\tau^2$. }}
\end{figure}

\begin{figure}[h]
\begin{center}
\includegraphics[width=0.49\linewidth]{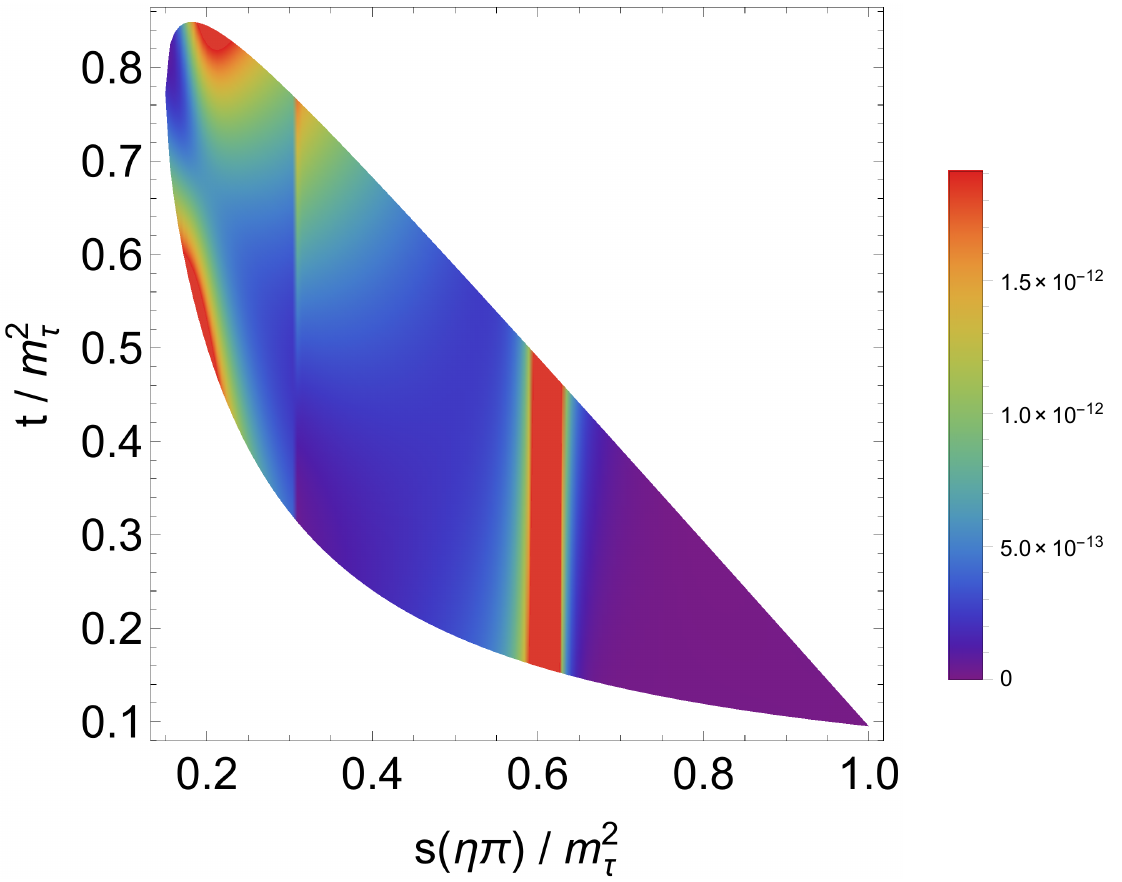} 
\includegraphics[width=0.49\linewidth]{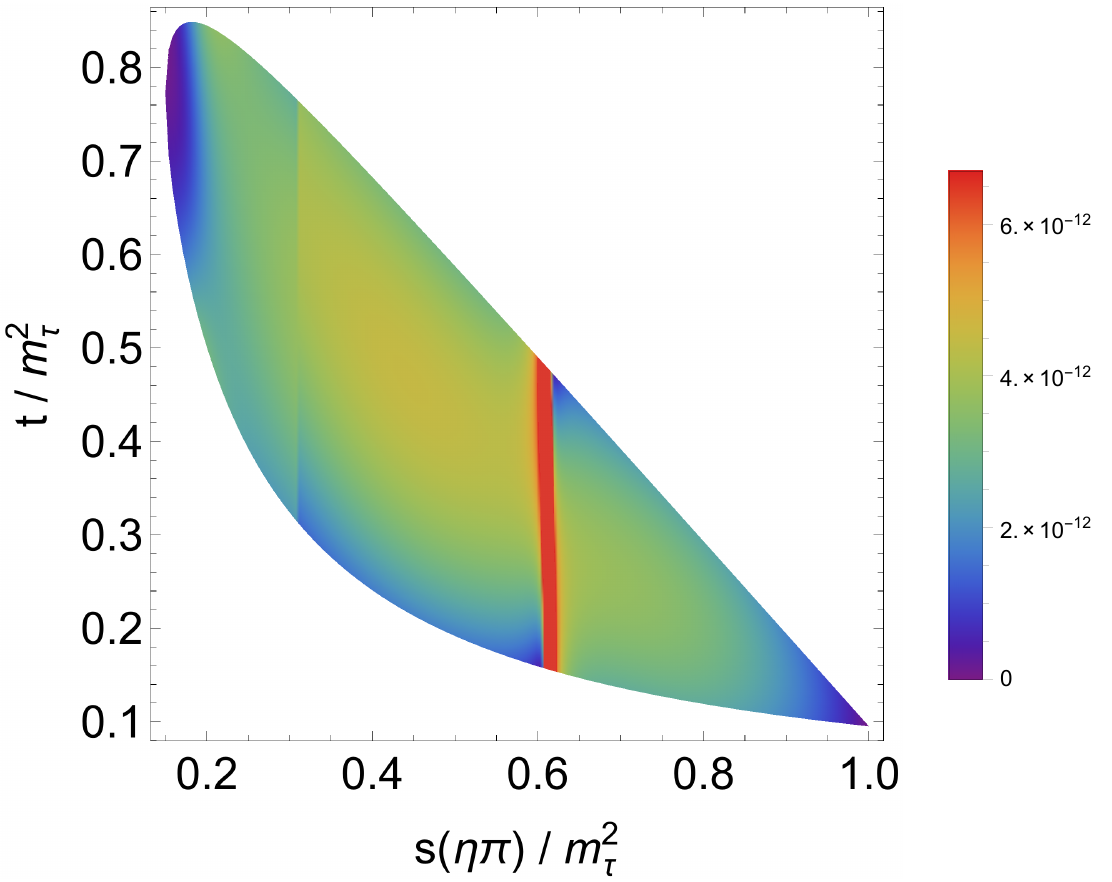} 
\includegraphics[width=.495\textwidth]{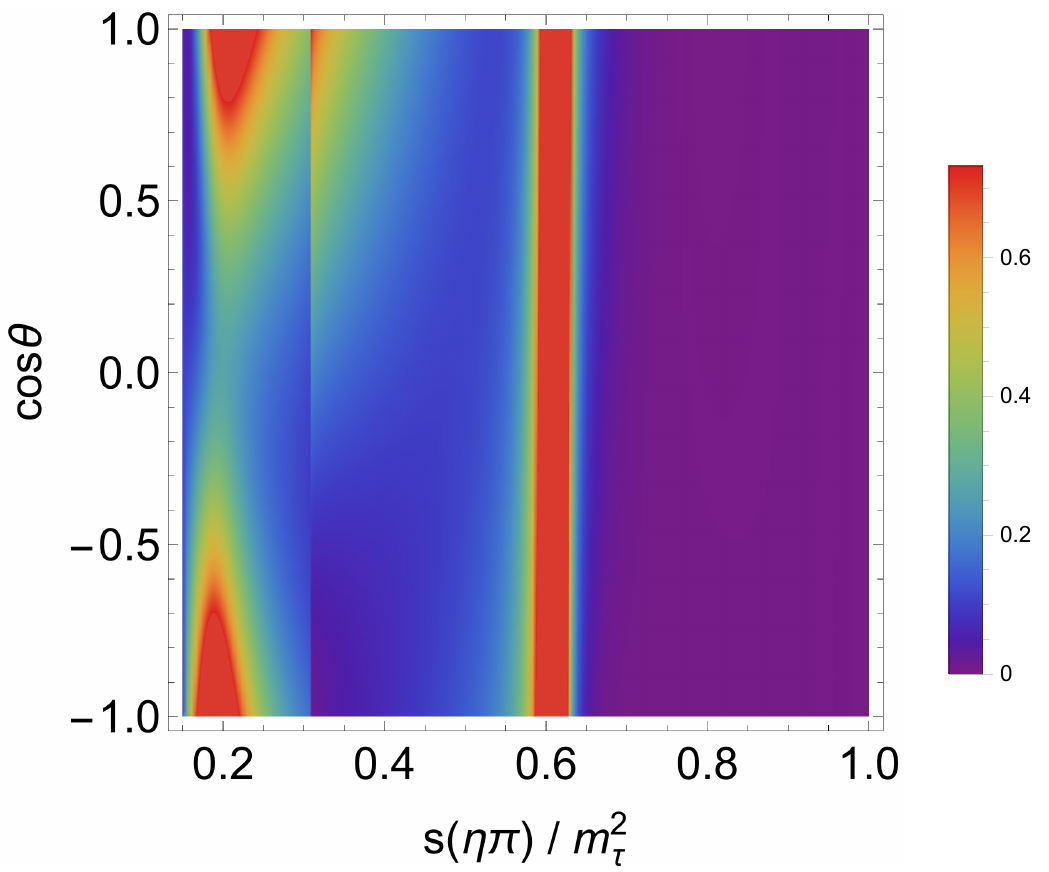}
\includegraphics[width=.495\textwidth]{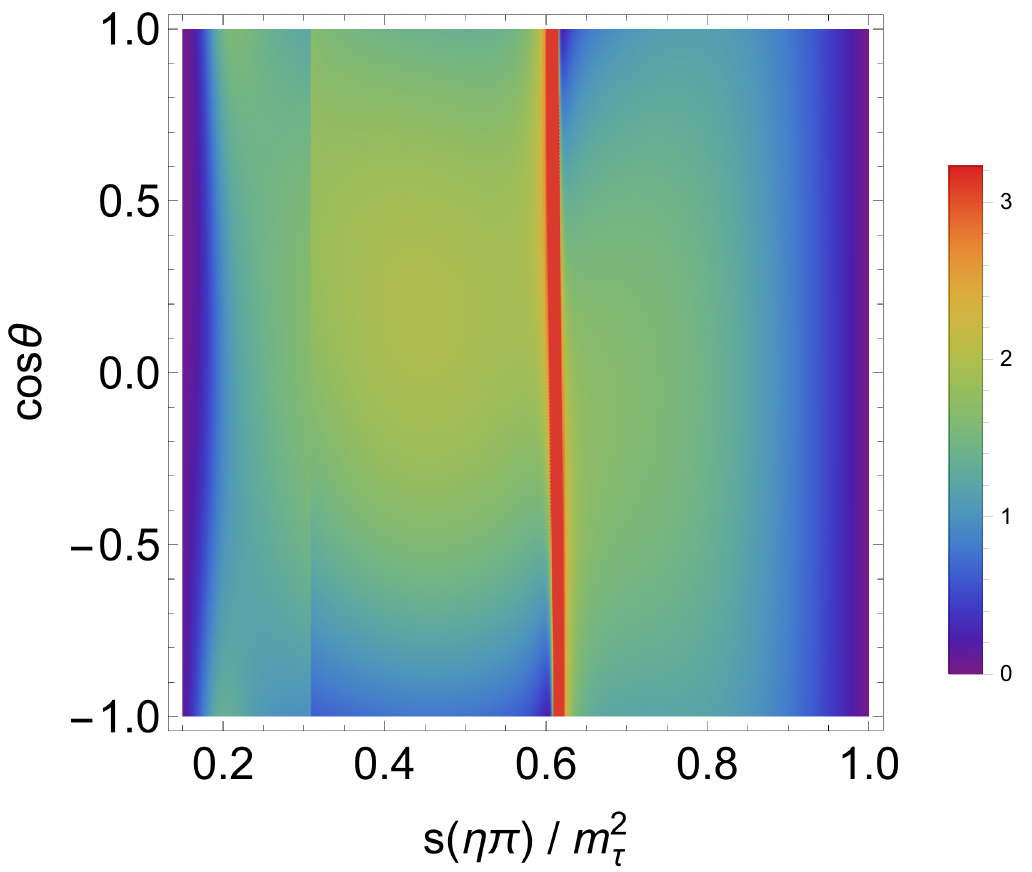}
\end{center}
    \caption{\label{fig:DalizM} Dalitz plot distribution for $\tau^- \to \pi^-\eta \nu_{\tau}$ decays: left-hand side corresponds to 
    $( \widehat{\epsilon}_S=0.002,\widehat{\epsilon}_{T}=0)$,
    while the figures in the right-hand side are obtained with the choice $(\widehat{\epsilon}_S=0,\widehat{\epsilon}_{T}=0.3)$. 
    The figures in the first row correspond to eq.~(\ref{M2}). Figures in the lower row corresponding to eq.~(\ref{angulardistribution}) are normalized to $\Gamma_\tau$. The Mandelstam variables, $s$ and $t$, 
    are normalized to $M_\tau^2$.}
\end{figure}

New Physics effects can appear in the distribution of Dalitz plots, with a large enhancement expected towards large values of the hadronic invariant mass (note eq. (\ref{s-dependence epsilon_S})). 
The first line of figure~\ref{fig:DalSM} shows the square of the matrix element $\overline{|{\cal M}|^2}_{00}$ obtained using the SM prediction for $\tau^- \to \pi^-\eta^{(\prime)} \nu_{\tau}$ 
form factors \cite{Escribano:2016ntp}; it can be appreciated that the dynamics is mainly driven by the scalar resonance with mass $\sim 1.39$ GeV (other two most populated spots in the Dalitz 
plot correspond to effects of the vector form factor, around the $\rho(770)$ peak, in the $\eta$ channel). In the first line of figure \ref{fig:DalizM} we show the squared matrix element $\overline{|{\cal M}|^2}$ 
for two representative values of the set of $(\widehat{\epsilon}_S, \widehat{\epsilon}_T)$ parameters that are consistent with current upper limits on the $B(\tau^- \to \pi^-\eta \nu_{\tau})$. A 
comparison of the plots in the first line of figure \ref{fig:DalSM} (left panel) and figures \ref{fig:DalizM} show that the Dalitz plot distribution is sensitive to the effects of tensor 
interactions but rather insensitive to the scalar interactions. For these, the most probable area around the $\rho$ peak gets thinner, while the one corresponding to the $a_0(1450)$ state 
gets wider, compared to the SM case. In the case of tensor interactions, the effect of the $\rho$ is diluted and the $a_0(1450)$ effect is also less marked than in the standard case. Given 
the fact that the $\rho$ contribution to these processes is much better known than that of the $a_0(1450)$, observing a weak $\rho$ meson effect in the Dalitz plot could be a signature of 
non-standard interactions, either of scalar or tensor type. Uncertainties on the scalar form factor prevent, at the moment, distinguishing between both new physics types by this Dalitz plot 
analyses.

\begin{figure}[h!] 
                \includegraphics[width=0.49\linewidth]{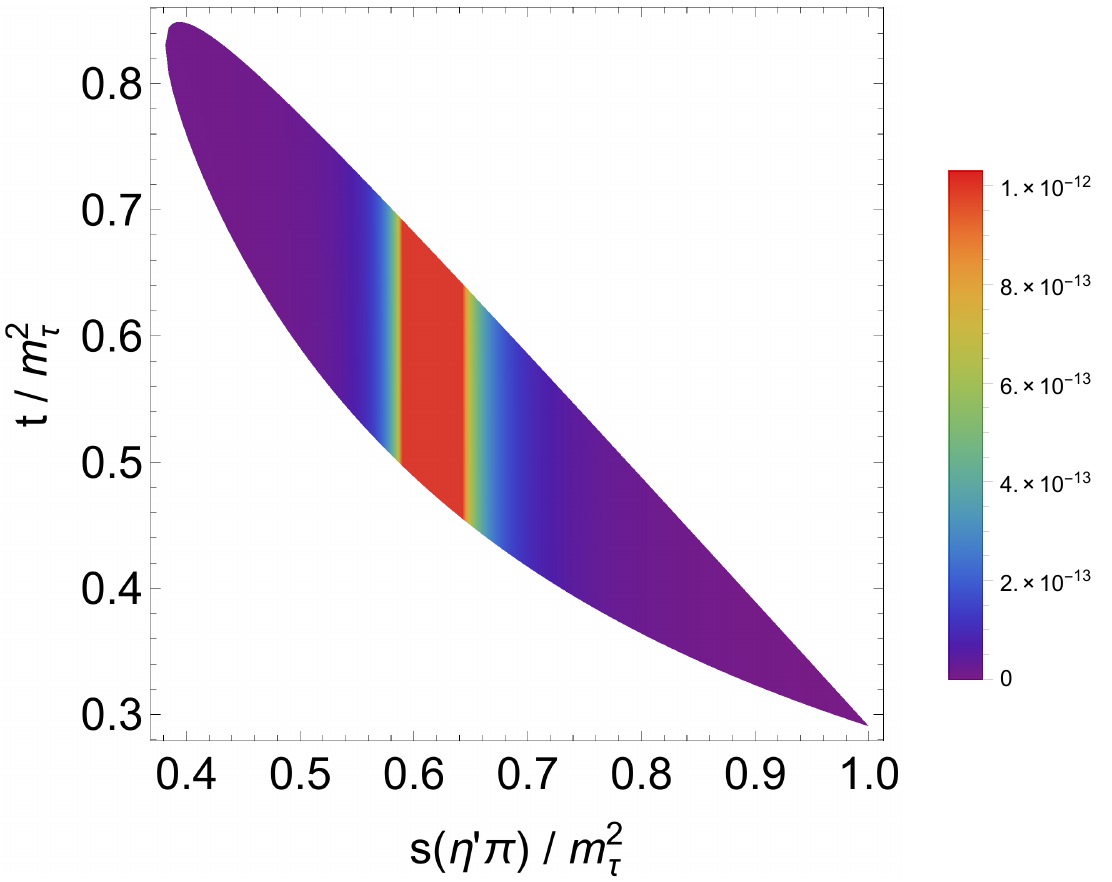} 
        \includegraphics[width=0.49\linewidth]{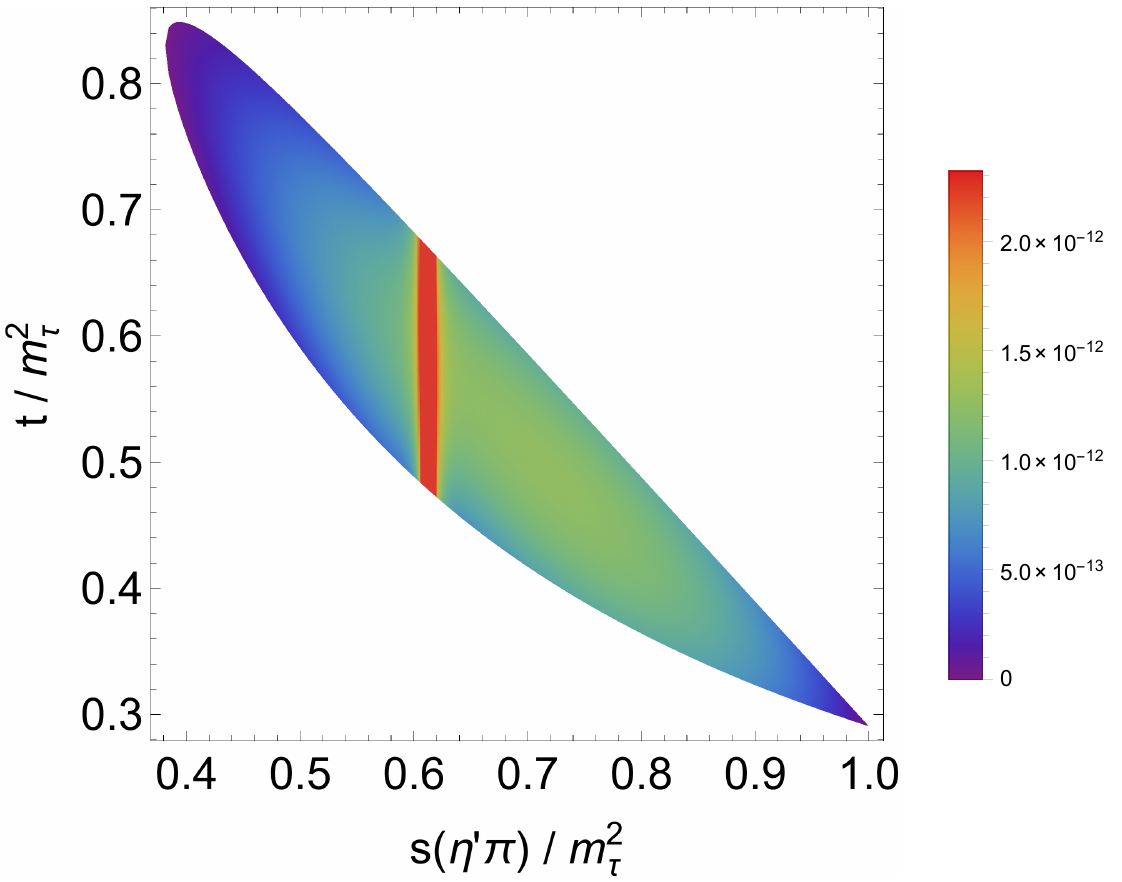} 
                \includegraphics[width=0.49\linewidth]{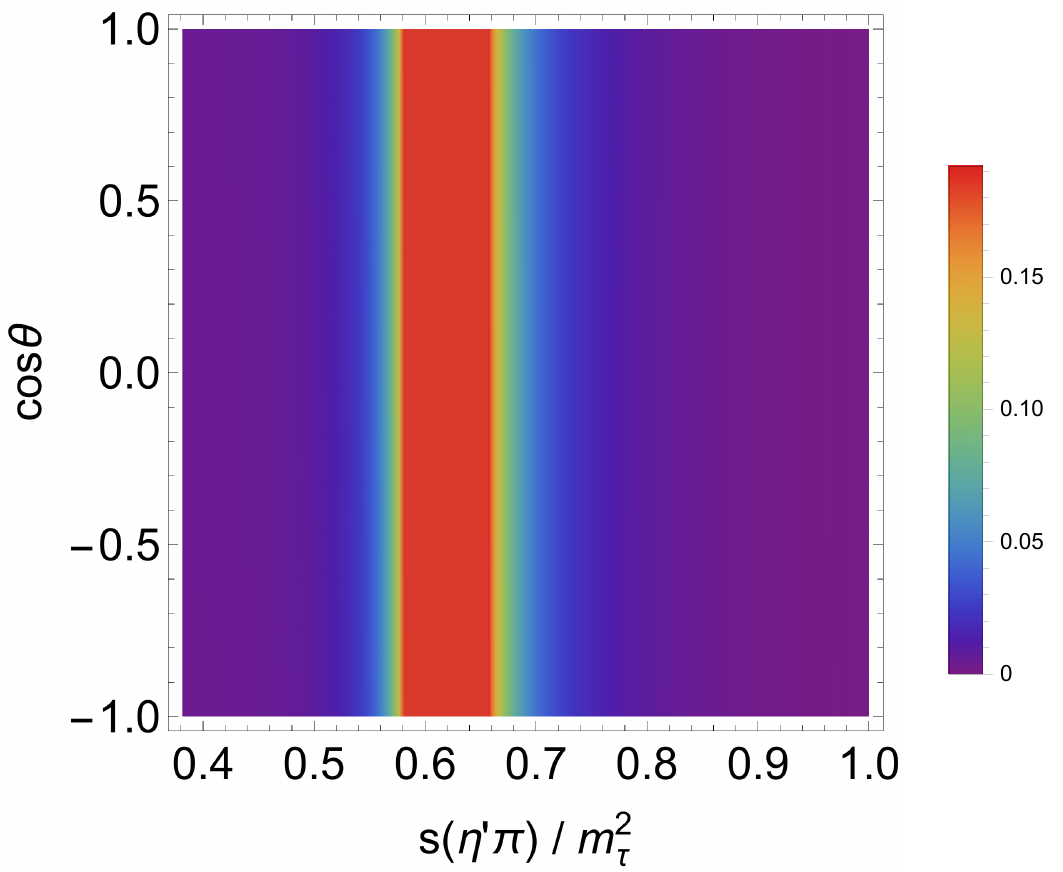} 
                \includegraphics[width=0.49\linewidth]{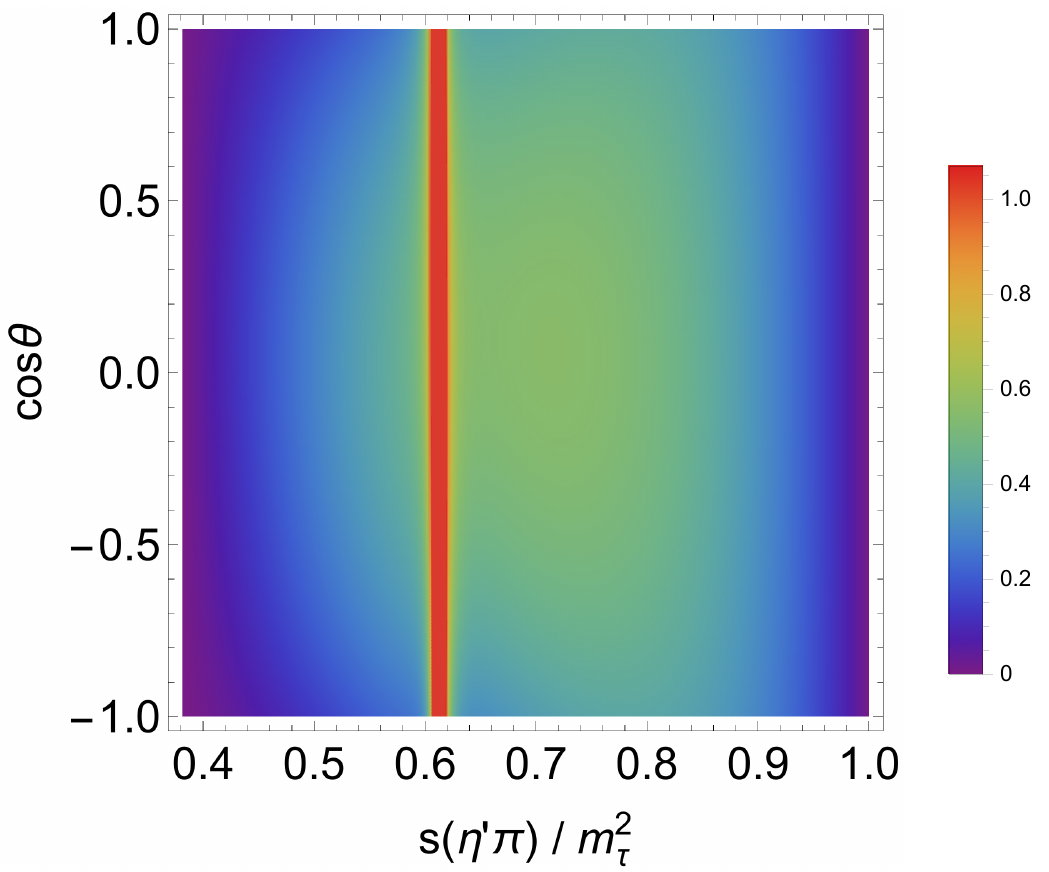} 
    \caption{\label{fig:DalizMP}Dalitz plot distribution for $\tau^- \to \pi^-\eta^{\prime} \nu_{\tau}$ decays: figures in the left correspond to $( \widehat{\epsilon}_S=0.006,\widehat{\epsilon}_{T}=0)$,
    while those in the right side are obtained with the choice $( \widehat{\epsilon}_S=0,\widehat{\epsilon}_{T}=10)$. The figures in the first row correspond to eq.~(\ref{M2}).
        The figures in the lower row to corresponding to eq.~(\ref{angulardistribution}) are normalized to the tau width. 
        The Mandelstam variables, $s$ and $t$, are normalized to $M_\tau^2$. }
\end{figure}

In the case of $\tau^- \to \pi^-\eta^{\prime} \nu_{\tau}$ decays the vector form factor contributes negligibly. Then, a comparison of the first rows of figures \ref{fig:DalSM} 
(right panel) and \ref{fig:DalizMP} (where the representative allowed values of $(\widehat{\epsilon}_S, \widehat{\epsilon}_T)$ differ from those taken for the $\eta$ channel) shows almost 
no change for scalar new physics. Tensor current contributions would decrease the $a_0(1450)$ effect compared to the SM. However, uncertainties on the scalar form factor will prevent 
drawing any strong conclusion from this feature.


\subsection{Angular distribution}
 
The hadronic mass and angular distributions of decay products are also modified by the effects of New Physics contributions and can offer a different sensitivity to the scalar and tensor 
interactions. For this purpose it becomes convenient to set in the rest frame of the hadronic system defined by 
 $\vec{p}_{\pi}+\vec{p}_{\eta^{(\prime)}}=\vec{p}_{\tau}-\vec{p}_{\nu_{\tau}}=0$. In this frame, the pion and tau lepton energies are given by 
 $E_{\tau}=(s+M_\tau^2)/2\sqrt{s}$ and $E_{\pi}=(s+m_{\pi}^2-m_{\eta^{(\prime)}}^2)/2\sqrt{s}$. The angle $\theta$ between the three-momenta of the pion and tau lepton is related to the invariant 
 $t$ variable by $t=m_{\tau}^2+m_{\pi}^2-2E_{\tau}E_{\pi}+2|\vec{p}_{\pi}| |\vec{p}_{\tau}|\cos \theta$, where $|\vec{p}_{\pi}|=\sqrt{E_{\pi}^2-m_{\pi}^2}$ and $|\vec{p}_{\tau}|=\sqrt{E_{\tau}^2-m_{\tau}^2}$.

The decay distribution in the $(s,\cos \theta)$ variables in the framework of the most general effective interactions is given by
\begin{eqnarray}\label{angulardistribution}
\frac{d^2 \Gamma}{d\sqrt{s}d\cos \theta} &=& \frac{ G_F^2 |V_{ud}|^2 S_{EW} }{128\pi^3 m_\tau} (1+\epsilon_L+\epsilon_R)^2 \left( \frac{m^2_{\tau}}{s}-1 \right)^2 |\vec{p}_{\pi}| 
\left\lbrace  (c_S\Delta^{\rm QCD}_{K^0K^+})^{2} |{F}_{0}^{\pi^-\eta^{(\prime)}}(s)|^2 \right.  \nonumber  \\
&& \left. \times \left(1 +\frac{ s\widehat{\epsilon}_S}{m_{\tau}(m_d-m_u)} \right)^2  +   16 |\vec{p}_{\pi}|^2 s^2  
\left| \frac{c_V}{2m_{\tau}}F_{+}^{\pi^-\eta^{(\prime)}}(s)-\widehat{\epsilon}_T F_T\right|^2
\right. \nonumber \\ 
&& \left.+ 4|\vec{p}_{\pi}|^2 s\left(1-\frac{s}{m^2_{\tau}} \right) \left[c^2_V|F_{+}^{\pi^-\eta^{(\prime)}}(s)|^2+4\widehat{\epsilon}_TF^2_T s \right]\cos^2 \theta + 4 c_S\Delta^{\rm QCD}_{K^0K^+} |\vec{p}_{\pi}| \sqrt{s} \cos\theta \right. \nonumber \\  
  && \left.  \ \times  \left(1 +\frac{ s\widehat{\epsilon}_S}{m_{\tau}(m_d-m_u)}\right) \left[c_V {\rm Re}[F_0(s) F_+^*(s)]-2\frac{s}{m_\tau}\widehat{\epsilon_T} F_T 
 {\rm Re}[F_0(s)]\right]\right\rbrace \ .
\end{eqnarray}

When the effective couplings of new interactions are turned off, we recover the usual expressions for this observable in the SM \cite{Beldjoudi:1994hi}. It is interesting to observe that no 
new angular dependencies appear owing to the presence of new interactions, although the coefficients of $\cos \theta$ terms get modified by terms that 
increase with the hadronic invariant mass $s$. {In this respect, it is interesting to point out that the last term of eq.~(\ref{angulardistribution}), 
which is linear in cos$\theta$, would allow to probe the relative phase between the scalar and vector contributions in the absence of new physics.}
We note that similar modifications to the angular and hadronic-mass distributions are expected for allowed $\tau^- \to (P_1P_2)^-\nu_{\tau}$ decays, although the effects of scalar and tensor 
interactions should be very small in those cases.

Results obtained using eq.(\ref{angulardistribution}) are plotted in the second row of figure \ref{fig:DalSM} for $\eta\pi^-$  ($\eta'\pi^-$) in the left (right) panel for the SM case. In the second row of figures 
\ref{fig:DalizM}, \ref{fig:DalizMP} we plot the 
$(s, \cos \theta)$ distributions,  which are defined from eq. (\ref{angulardistribution}),
using the same representative values of $(\widehat{\epsilon}_S, \widehat{\epsilon}_T)$ parameters for every channel employed above.

In general, a comparison between figures \ref{fig:DalSM}, \ref{fig:DalizM} and \ref{fig:DalizMP} shows that, remarkably, differences between SM and New Physics distributions can be obtained either using the  $(s,t)$ or the $(s,$ cos$\theta)$ Dalitz plot analyses. Then, the experimentally cleanest of these will be more useful restricting non-standard interactions. If both are available, consistency checks can be done  by comparing their respective data.

\subsection{Decay rate}
Integration upon the $t$ variable in eq. (\ref{dalitz}) gives the hadronic invariant mass distributions 

\begin{eqnarray} \label{had-dist}
\frac{d\Gamma}{ds} &=&\frac{G_F^2 S_{EW} m_{\tau }^3 \Big|V_{ud}F_+^{\pi^-\eta^{(\prime)}}(0)\Big|^2}{384\pi^3 s} (1+ \epsilon_L+\epsilon_R )^2   \left(1 -\frac{s}{ m_{\tau}^2}\right)^2   \lambda^{1/2}\left({s,m_{\eta^{(\prime)}} }^2,{m_\pi }^2\right)  \nonumber \\ 
&& \ \ \ \ \ \ \ \ \ \ \ \ \ \times  \left[ X_{VA}+\widehat{\epsilon}_S X_{S}+ 
\widehat{\epsilon}_{T} X_{T} + \widehat{\epsilon}_S^2 X_{S^2}+ \widehat{ \epsilon}_{T}^2 X_{T^2}   
\right] ,\,
\end{eqnarray}
\normalsize
where 
\begin{eqnarray}
X_{VA} &= &\frac{1}{s^{2}}
\left[ 3 |\widetilde{F}_{0}^{\pi^-\eta^{(\prime)}}(s)|^2 \Delta_{\pi^-\eta^{(\prime)}}^2  + |\widetilde{F}_{+}^{\pi^-\eta^{(\prime)}}(s)|^2 \lambda \left(s, {m_{\eta^{(\prime)}} }^2, {m_\pi }^2\right) \left(1+\frac{2s}{m_\tau^2}\right)  \right]\,, \nonumber \\
X_{S} &=& \frac{6}{ s~ m_\tau } 
  |\widetilde{F}_{0}^{\pi^-\eta^{(\prime)}}(s)|^2  ] \frac{\Delta_{\pi^-\eta^{(\prime)}}^2 }{m_d-m_u}\,, \nonumber \\ 
X_{T}  &=& \frac{-6\sqrt{2}}{s ~m_\tau}  \frac{ \rm{Re} [F_+(s)] F_T}{|F_+^{\pi^-\eta^{(\prime)}}(0)|^2} \lambda \left(s, {m_{\eta^{(\prime)}} }^2, {m_\pi }^2\right) \,, \nonumber \\
X_{S^2} &= & \frac{3}{m_\tau^2} 
 |\widetilde{F}_{0}^{\pi^-\eta^{(\prime)}}(s))|^2 \frac{ \Delta_{\pi^-\eta^{(\prime)}}^2}{(m_d-m_u)^2} \,, \\ \nonumber
X_{T^2}  &= &\frac{4}{s} \frac{ |{F_T}|^2}{|F_+(0)|^2}  \left(1 + \frac{s}{2m_\tau^2} \right) \lambda \left(s, {m_\eta^{(\prime)} }^2, {m_\pi }^2\right)\,. \\ \nonumber
\end{eqnarray}
\normalsize
Notice that when $\epsilon_L = \epsilon_R = \widehat{\epsilon}_S= \widehat{\epsilon}_T=0$ we recover the SM result from \cite{Escribano:2016ntp}. 
We also note that by using finiteness of the matrix element at the origin, and the fact that the form 
factors are normalized at the origin, we have \cite{Escribano:2016ntp}
\be
F_+^{\pi^-\eta^{(\prime)}}(0)=
-\frac{c^S_{\pi^-\eta^{(\prime)}}}{c^V_{\pi^-\eta^{(\prime)}}}
\frac{\Delta^{\rm QCD}_{K^0K^+}}{\Delta_{\pi^-\eta^{(\prime)}}}F_0^{\pi^-\eta^{(\prime)}}(0)\ ,
\label{VFF0} 
\ee
and
\be
\widetilde{F}_{+,0}^{\pi^-\eta^{(\prime)}}(s)=\frac{F_{+,0}^{\pi^-\eta^{(\prime)}}(s)}{F_{+,0}^{\pi^-\eta^{(\prime)}}(0)}\ ,
\ee
which have been used to write eq.~(\ref{had-dist}).

\begin{figure}[t!]
        \includegraphics[width=0.49\linewidth]{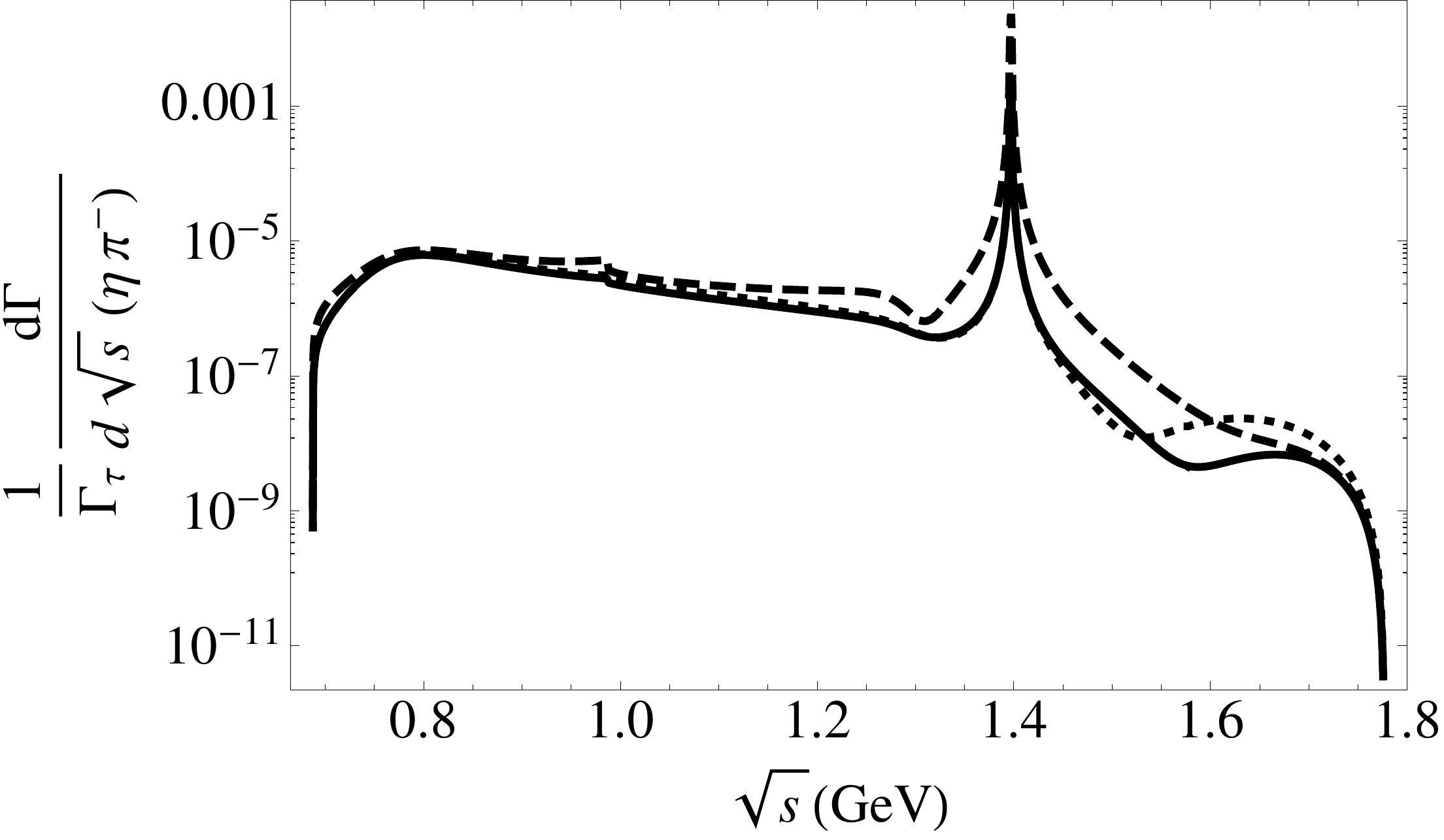}
        \includegraphics[width=0.49\linewidth]{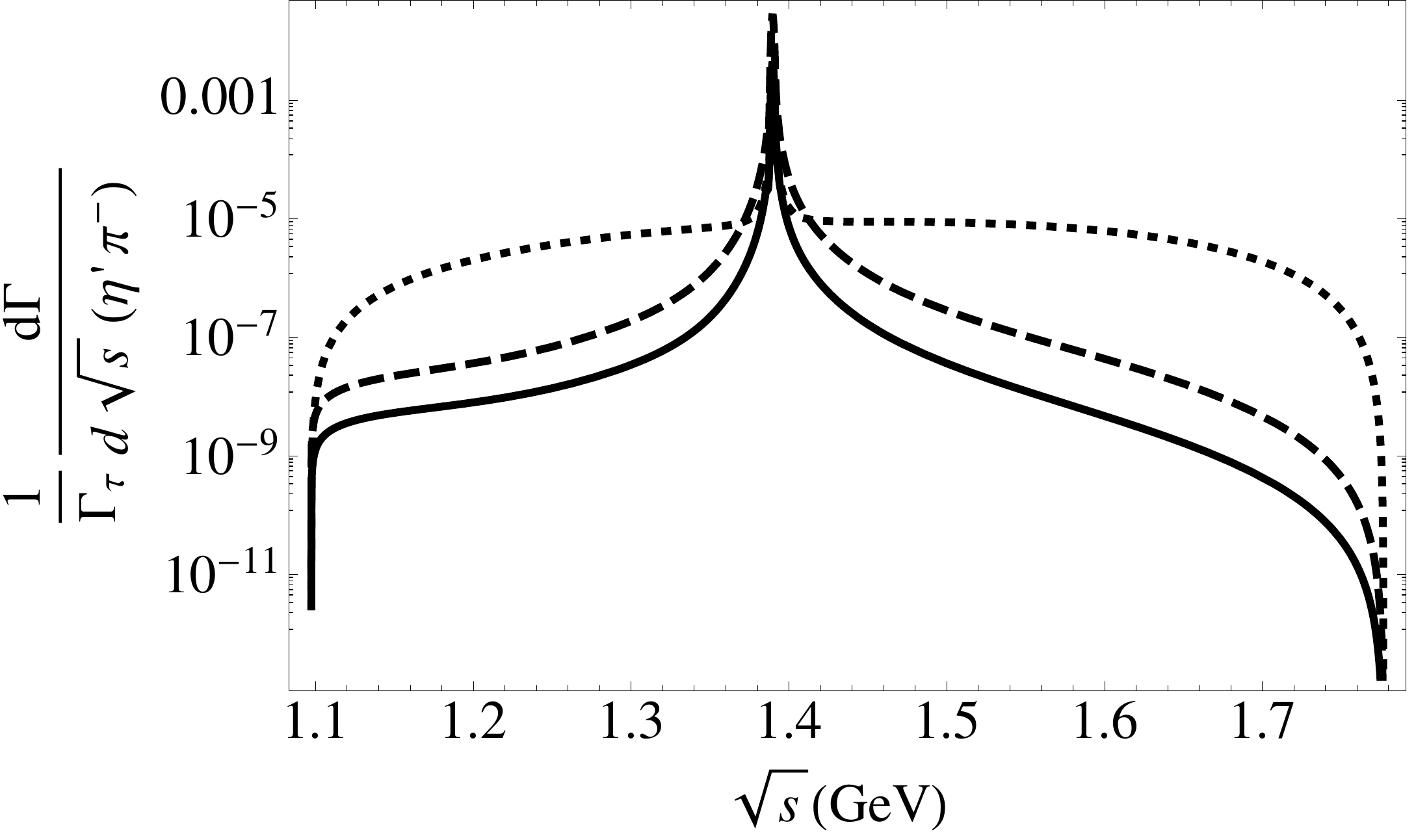} 
\caption{ \label{Fig:Spec} \small Left figure shows the $\eta\pi^-$  hadronic invariant mass distribution for the SM (solid line) and $\widehat{\epsilon_s}=0.004,\widehat{\epsilon_T}=0$ (dashed line), 
$\widehat{\epsilon_s}=0,\;\widehat{\epsilon_T}=0.6$ (dotted line). Right figure shows the $\eta'\pi^-$ hadronic invariant mass distribution for the SM (solid line) and 
$\widehat{\epsilon_s}=0.005,\;\widehat{\epsilon_T}=0$ (dashed line), $\widehat{\epsilon_s}=0,\widehat{\epsilon_T}=10$ (dotted line). Units in axis are given in powers of GeV 
{and the decay distributions are normalized to the tau decay width}. }
\end{figure}

 In figure \ref{Fig:Spec} we plot the invariant mass distributions of the hadronic system for $\tau^- \to \pi^-\eta^{(\prime)}\nu_{\tau}$ decays. 
Noticeable differences are observed outside the resonance peak region ($M_S\sim1.39$ GeV, \cite{Escribano:2016ntp}) when we allow for small departures from the SM. Again, the hadronic spectrum in 
both cases ($\pi\eta$ and $\pi\eta'$) is able to distinguish New Physics contributions provided the scalar form factor contributions are known to a sufficient level of accuracy (we will 
quantify this statement in the next section). While the scalar non-standard interactions basically modify the spectrum (which essentially keeps its shape) as a global factor, tensor 
interactions act quite smoothly over the phase space (contrary to the scalar form factors, which are extremely peaked around $\sqrt{s}\sim1.39$ GeV). This would soften the $\eta$ channel 
spectrum visibly (in logarithmic scale). Since the $\eta'$ channel is so much dominated by the scalar form factor, the change in the spectrum would be even harder to be appreciated, 
and only a precise measurement of its tale could show a deviation from the SM case hinting to vector-tensor interference.

\section{Results and discussion}\label{Concl}

\begin{figure}[h!]
        \includegraphics[width=0.49\linewidth]{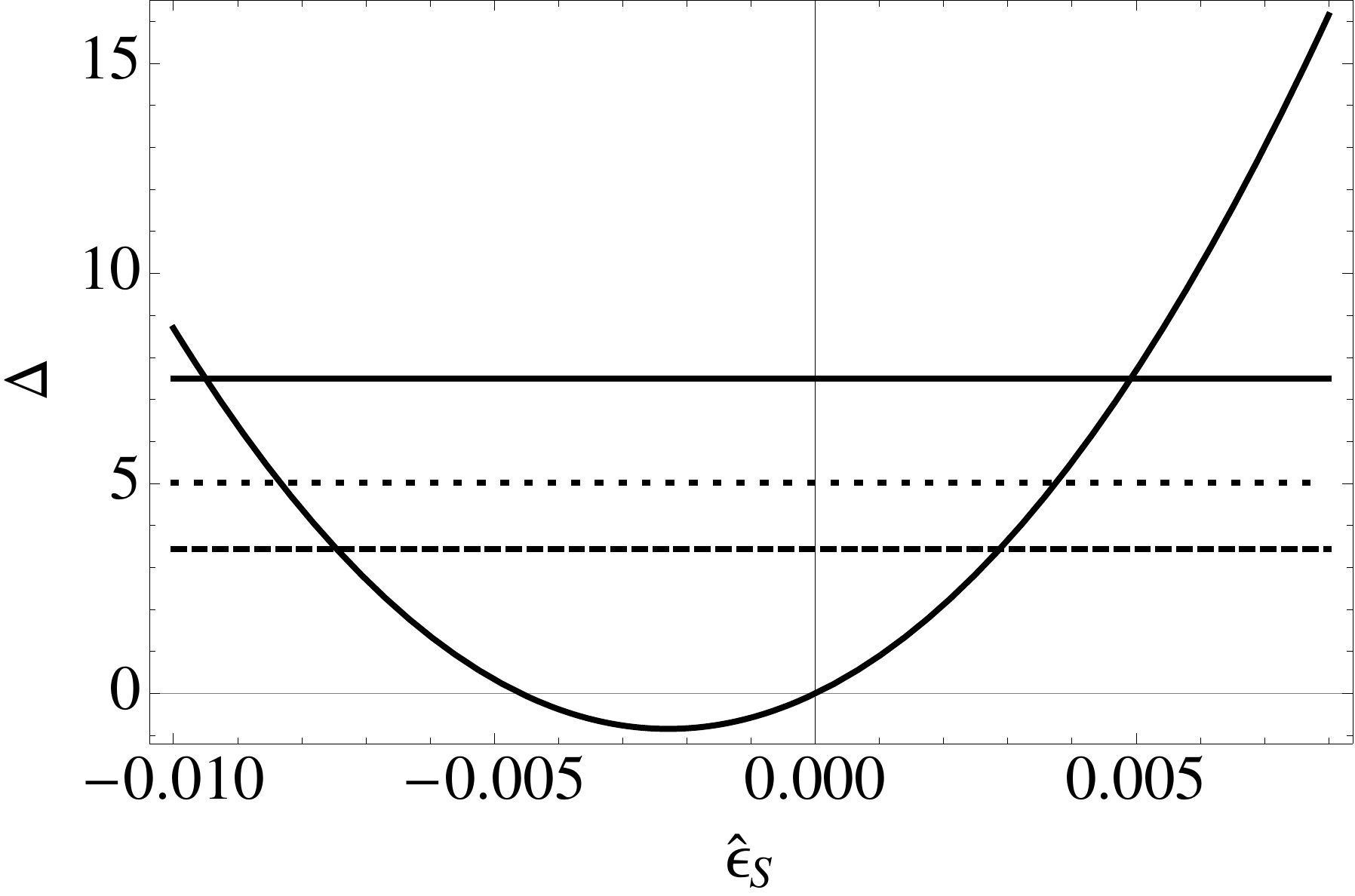} 
        \includegraphics[width=0.49\linewidth]{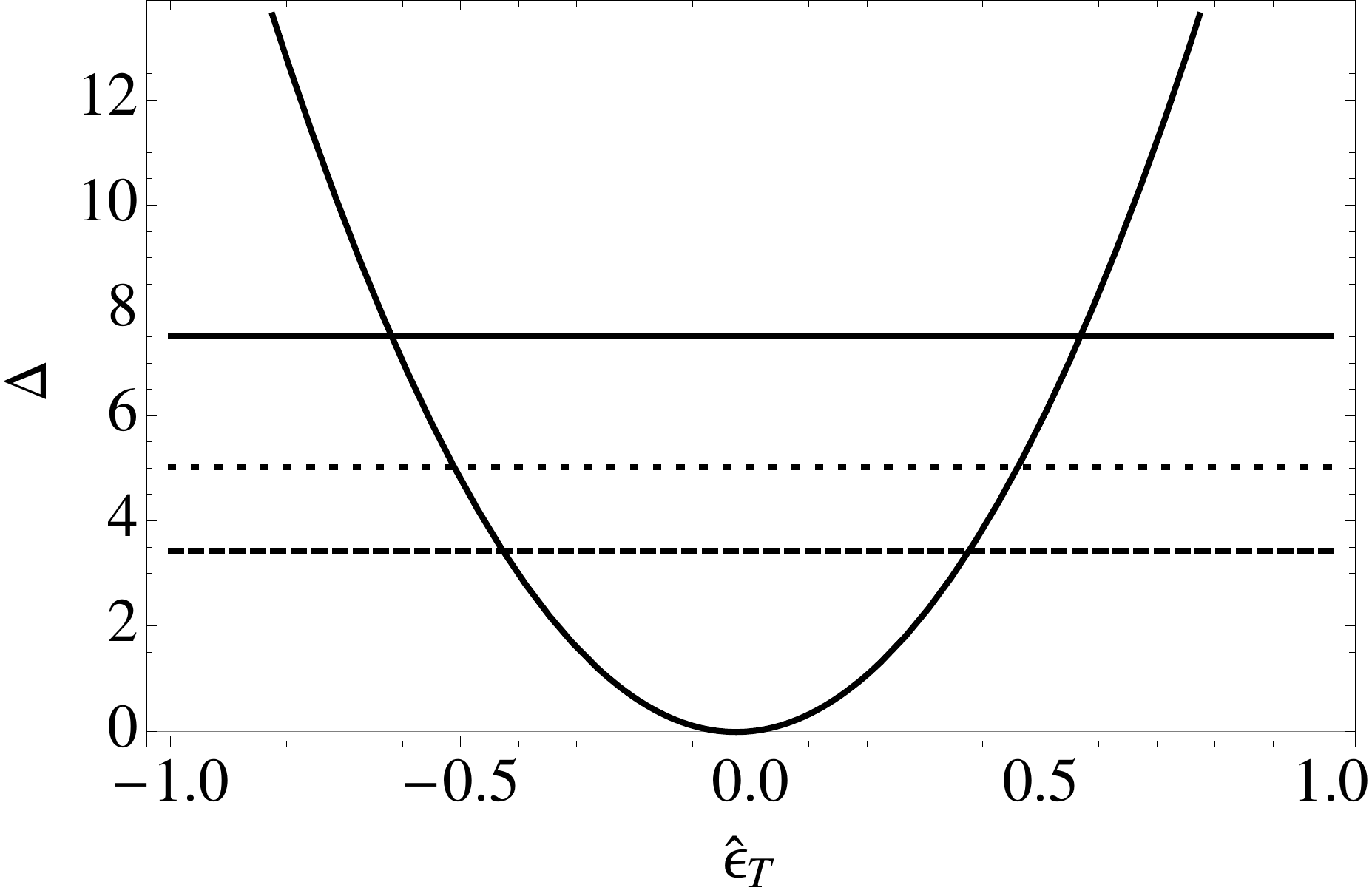} 
  \caption{\label{fig:Del}$\Delta$ as a function of $\widehat{\epsilon}_S$ (for $\widehat{\epsilon}_T=0)$ and $\widehat{\epsilon}_T$ (for $\widehat{\epsilon}_S=0)$ for 
    $\tau^-\to \pi^-\eta\nu_{\tau}$ decays. Horizontal lines represent current values of $\Delta$ according to the upper limits on the branching fraction obtained by Babar (dotted line), 
    $< 9.9\times 10^{-5}$, $95\%$ CL \cite{delAmoSanchez:2010pc}, Belle (dashed line), $< 7.3\times 10^{-5}$, $90\%$ CL \cite{Hayasaka:2009zz} and CLEO (solid line), 
    $< 1.4\times 10^{-4}$, $95\%$ CL \cite{Bartelt:1996iv}.}
\end{figure}
\begin{figure}[h!]
        \includegraphics[width=0.49\linewidth]{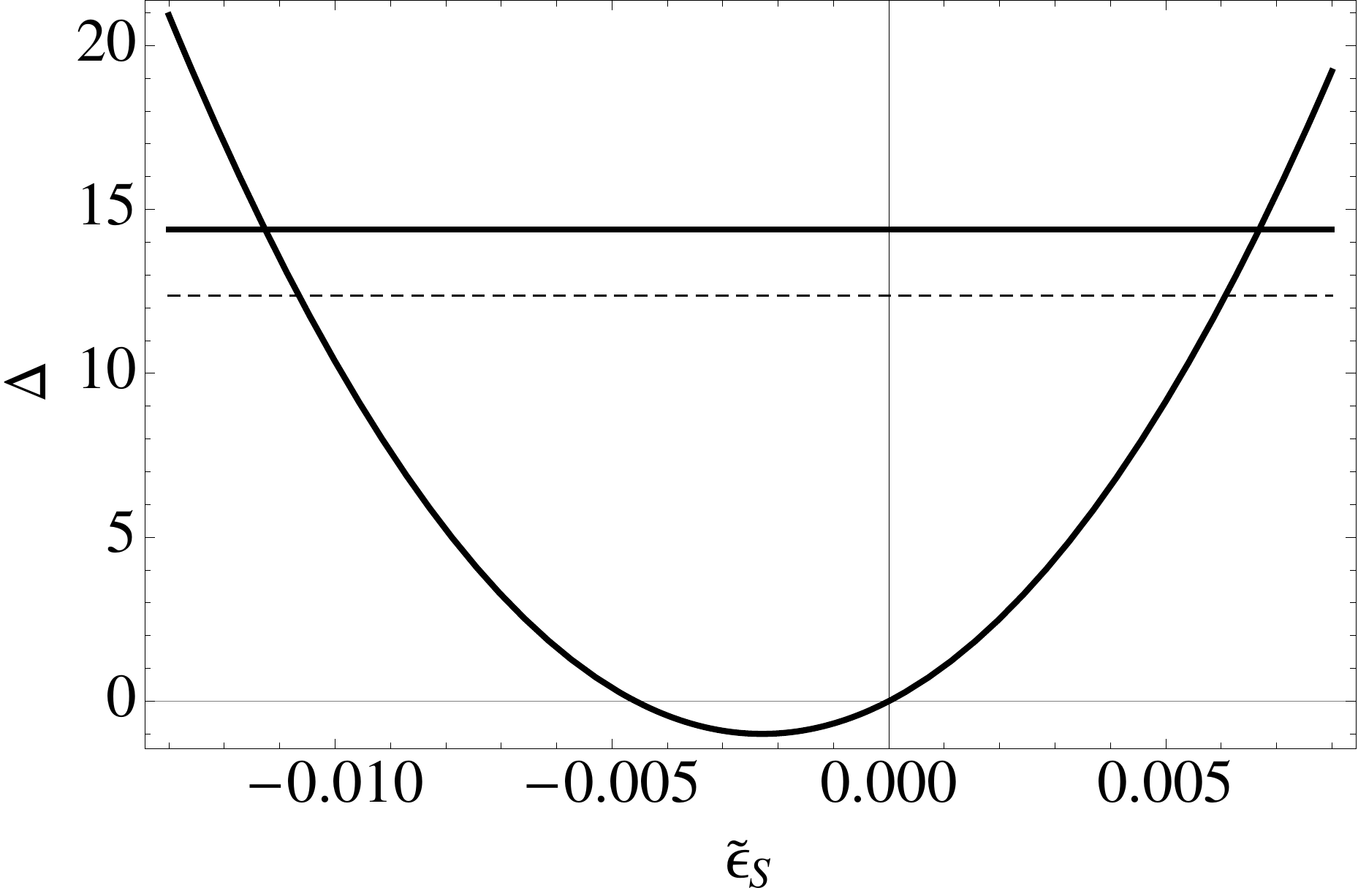} 
        \includegraphics[width=0.49\linewidth]{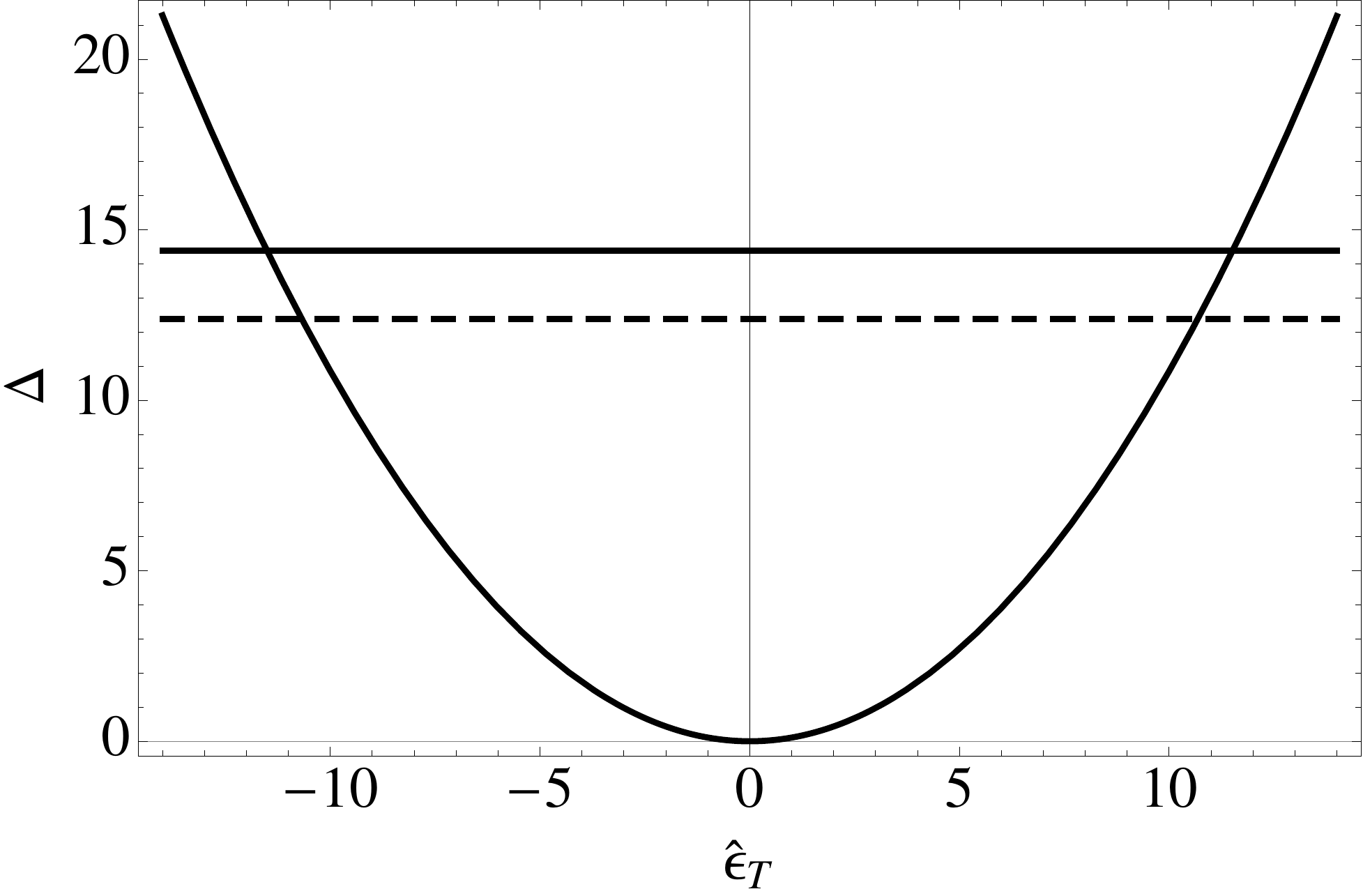} 
 \caption{\label{fig:Delp} $\Delta$ as a function of $\widehat{\epsilon}_S$ (for $\widehat{\epsilon}_T=0)$ and $\widehat{\epsilon}_T$ (for $\widehat{\epsilon}_S=0)$ for 
    $\tau^-\to \pi^-\eta'\nu_{\tau}$ decays. Horizontal lines represent current values of $\Delta$ according to the upper limits on the branching fraction obtained by Babar (solid line), 
    $<7.2\cdot 10^{-6}$, $95\%$ CL \cite{Aubert:2008nj} and Belle (dashed line),  $<4.6\cdot 10^{-6}$, $90\%$ CL \cite{Hayasaka:2009zz}.}
\end{figure}
\begin{figure}[h!]
\includegraphics[width=0.49\linewidth]{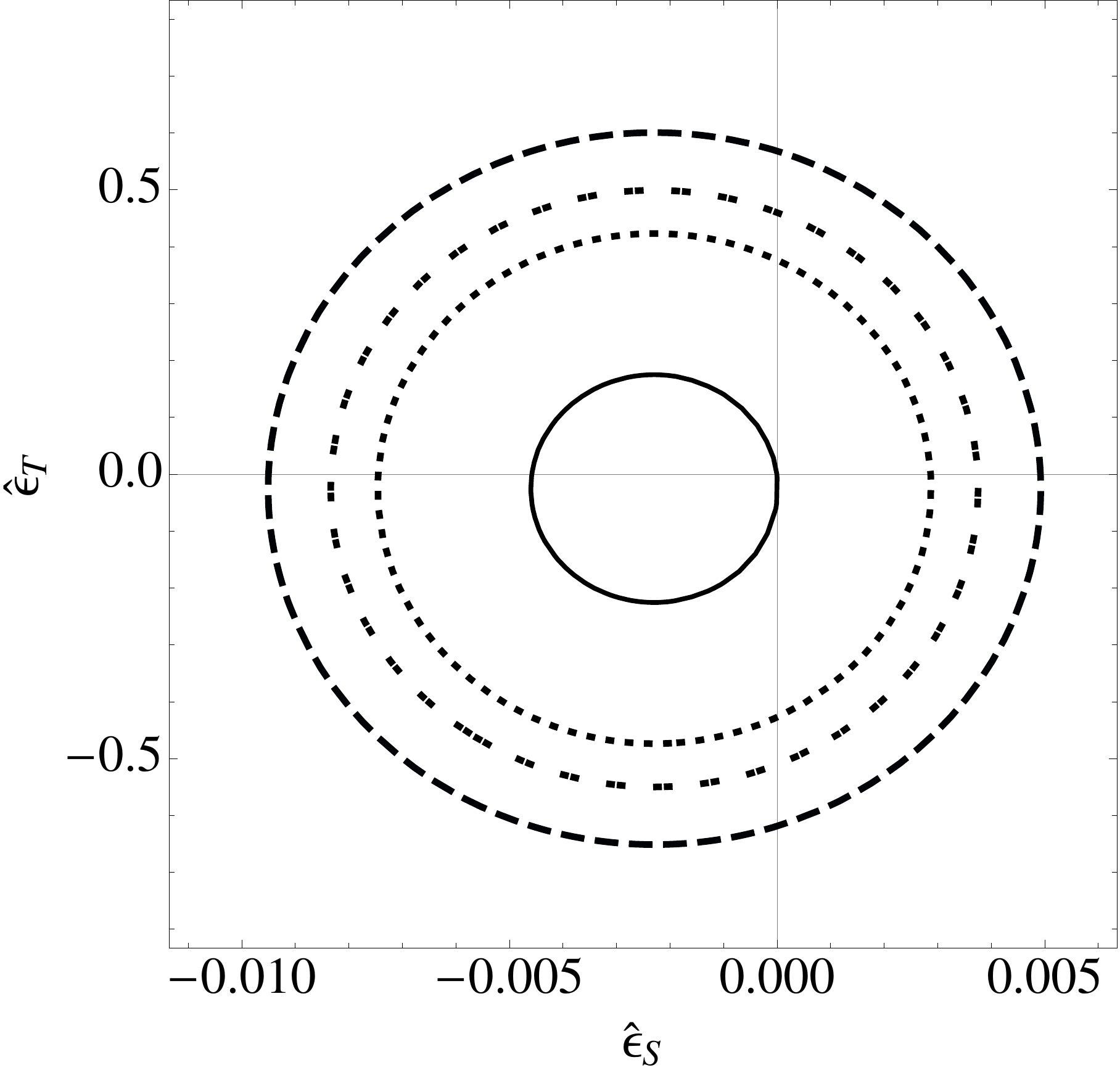}
        \includegraphics[width=0.49\linewidth]{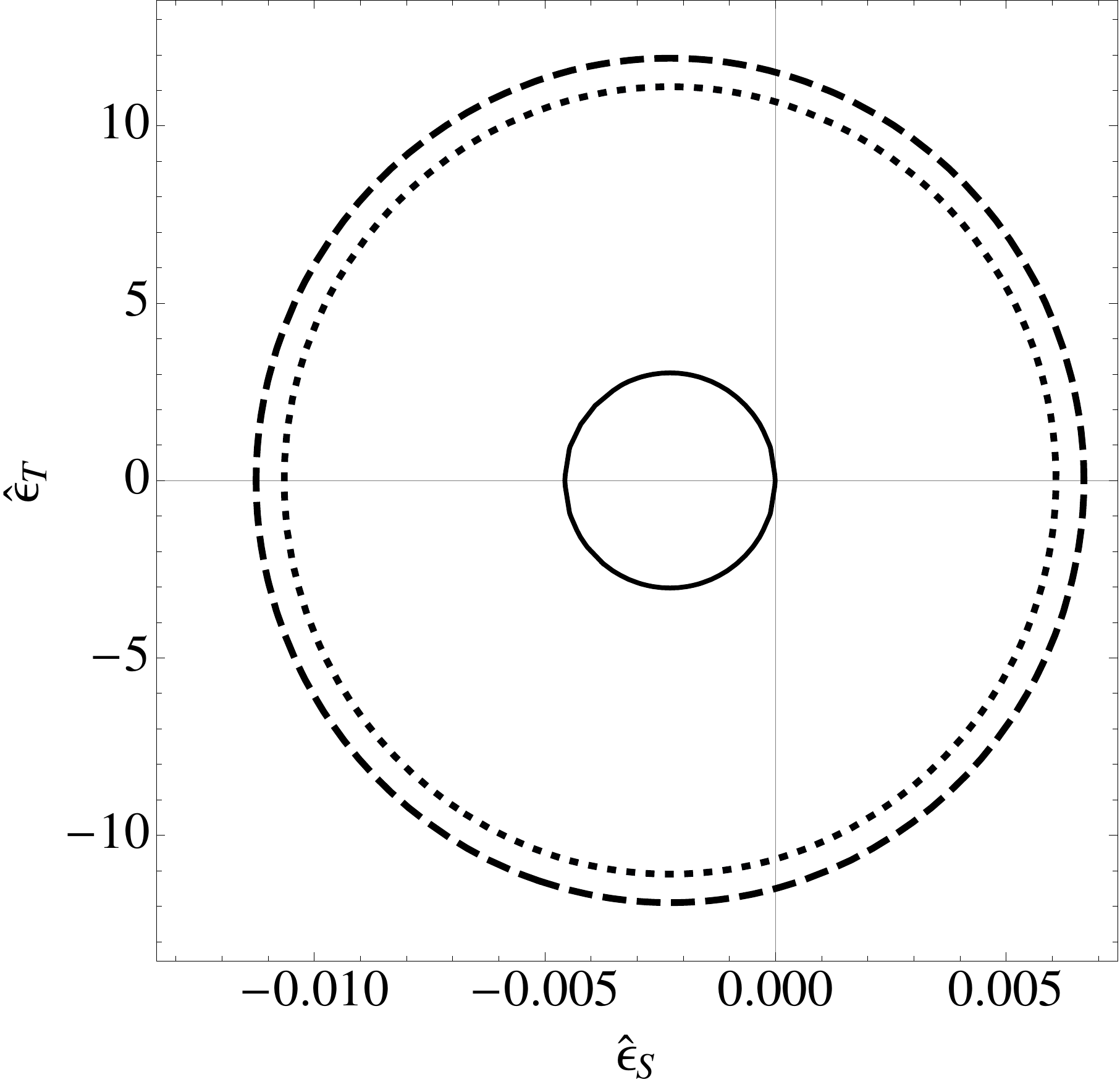} 
\caption{\label{fig:DelC}  Left figure shows constraints on scalar and tensor couplings obtained from $\Delta(\tau^-\to\eta\pi^-\nu_\tau)$ values using current experimental upper limits 
on branching fractions. The solid line represents $\Delta=0$, the dotted line is the Belle 90\%CL limit,  the doubly dotted line is  the BaBar 95\%CL limit and the dashed line is the 
CLEO 95\%CL limit. In the right side we have contours of constant $\Delta(\tau^-\to\eta'\pi^-\nu_\tau)$ in the $\widehat{\epsilon}_S-\widehat{\epsilon}_T$ plane. The inner solid circle is 
the SM prediction, $\Delta=0$, the dotted line is the BaBar 95\%CL limit and the dashed line is the Belle 90\%CL limit.}
\end{figure}
\begin{figure}[h!]
\includegraphics[width=0.49\linewidth]{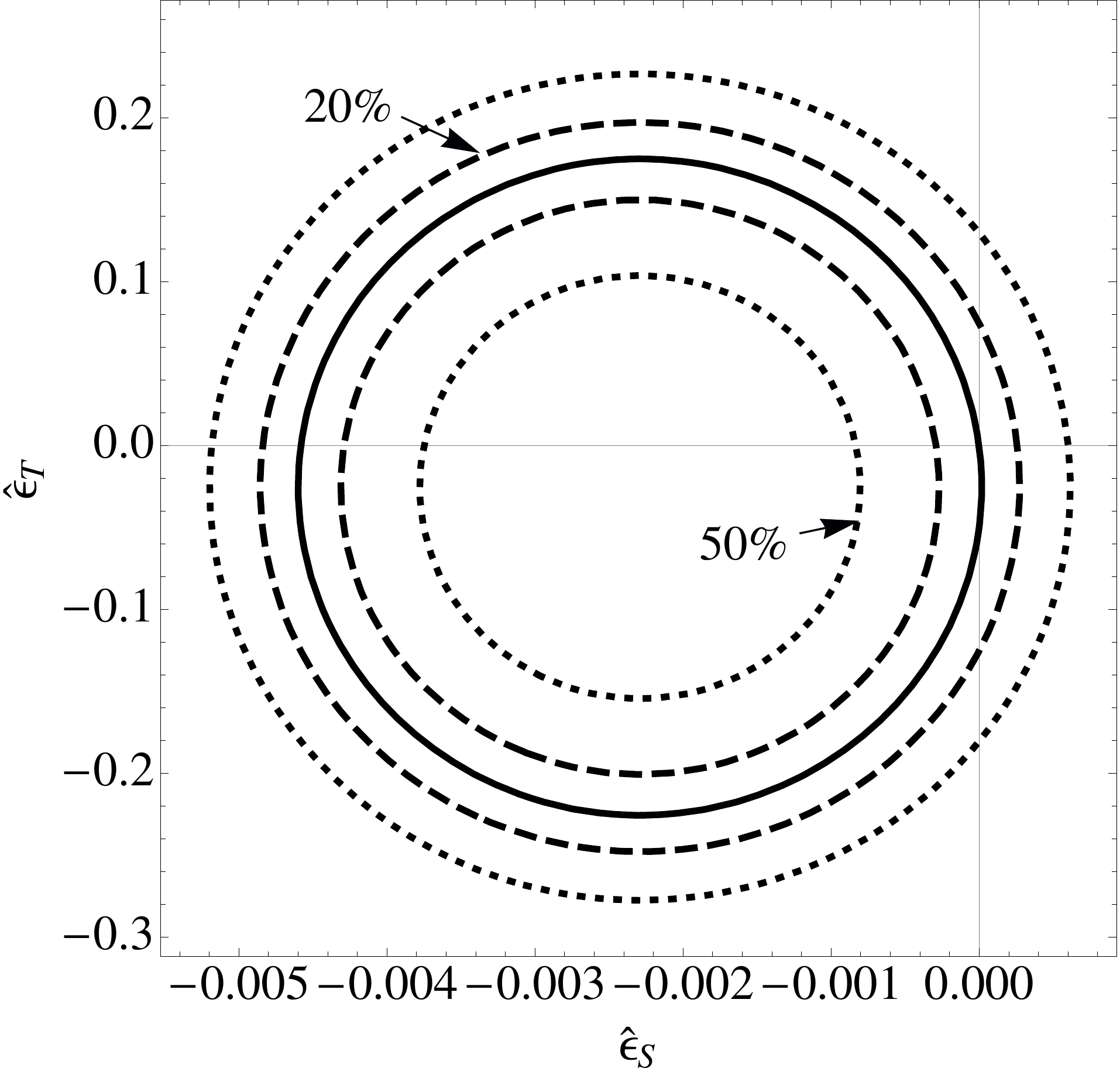}
        \includegraphics[width=0.49\linewidth]{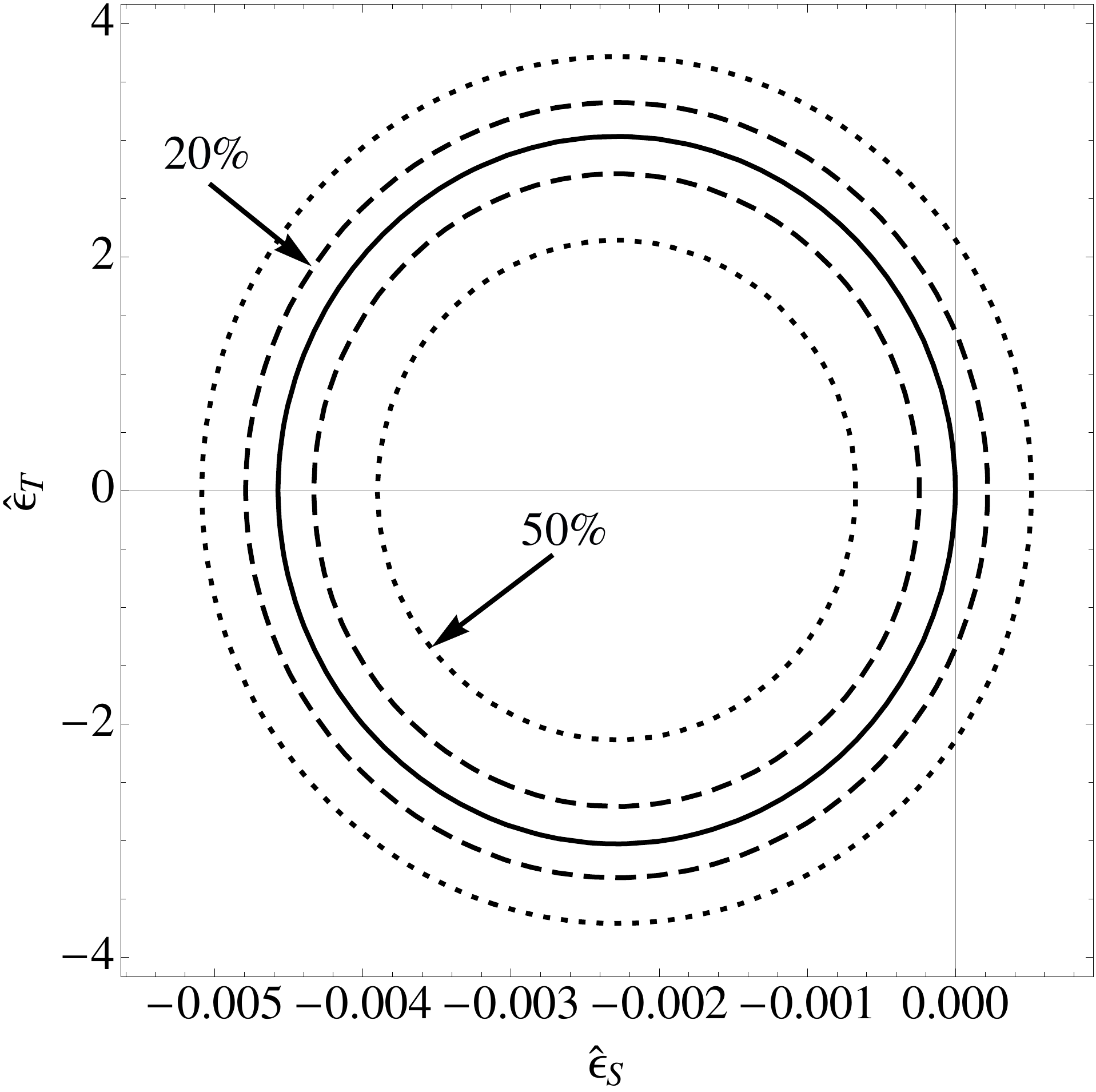} 
\caption{\label{fig:DelPresc} 
The thick solid line in the middle shows the contour for the SM Branching Ratio of $\tau^- \to \pi^-\eta^{(\prime)}\nu_{\tau}$. In the hypothetical case of this value being measured by 
Belle-II with 50\%(band bounded with dotted lines) and 20\%(band bounded with dashed lines) precision, some nonzero allowed range of values for $\widehat{\epsilon}_S,\widehat{\epsilon}_T$ can be determined. In the 
left-hand side we have the case of $\tau^- \to \pi^- \eta \nu_{\tau}$, while on the right $\tau^- \to \pi^- \eta^{\prime} \nu_{\tau}$ is shown.}
\end{figure}

 Equation (\ref{had-dist}) can be integrated to obtain the total decay rate of the $\tau^- \to \pi^-\eta^{(\prime)}\nu_{\tau}$ decays, using the expressions for the form factors discussed in Ref. \cite{Escribano:2016ntp} and in Section \ref{tensor}. Since the total 
 decay rate depends upon several effective couplings, we can explore how New physics effects inducing scalar and tensor interactions can be constrained by measurements of the branching fractions. 
 For this purpose, we compare the decay rate ($\Gamma$) for $\tau^- \to \pi^-\eta^{(\prime)}\nu_{\tau}$ including all the interactions with respect to the one ($\Gamma^0$) obtained by neglecting 
 $\widehat{\epsilon}_S$ and $\widehat{\epsilon}_T$ couplings. Integrating eq. (\ref{had-dist}) we get the shift produced by new physics contributions as follows
\begin{equation} \label{percent}
\Delta\equiv \frac{\Gamma-\Gamma^0}{\Gamma^0}=\alpha \widehat{\epsilon}_S  +
\beta \widehat{\epsilon}_T + \gamma \widehat{\epsilon}_S^2  + \delta \widehat{\epsilon}_T^2  \, .
\end{equation}
Clearly, $\Delta=0$ when we have only vector current contributions to the decay amplitude. The numerical values of the coefficients are: 
$\alpha\sim (7\cdot10^2, 9\cdot10^2) ,~ \beta\sim(1.1, -8\cdot 10^{-4}),~ \gamma\sim(1.6\cdot 10^5,1.9\cdot 10^5)$ and $\delta\sim(21, 0.1)$ 
where the first (second) value refers to $\pi\eta\; (\pi\eta^{\prime})$ channel. Easy-to-estimate uncertainties on these values are given by the corresponding errors of 
$\epsilon_{\pi\eta^{(\prime)}}$, given the quadratic dependence of observables on these mixing coefficients. {For the most interesting case of $\alpha^{\pi\eta}$, this 
yields the range $\left[300, 800\right]$, approximately.}

Eq. (\ref{percent}) is a quadratic function of the effective scalar and tensor couplings that can be used to explore the sensitivity of $\tau^- \to \pi^-\eta^{(\prime)}\nu_{\tau}$ decays to the 
effects of New Physics. This can be achieved in two different ways. Firstly, we can represent the constraint on scalar (tensor) couplings obtained from the current upper limits on $\Gamma$ by assuming 
$\widehat{\epsilon}_T=0$~(respectively, $\widehat{\epsilon}_S=0)$. This is shown in  figure~\ref{fig:Del} where we represent with horizontal lines the current experimental upper limits on $\Delta$ and eq. 
(\ref{percent}) for $\tau^- \to \pi^-\eta\nu_{\tau}$ decays. According to this procedure, we get the constraint $-0.008\leq \widehat{\epsilon}_S \leq 0.004$ which corresponds to the 
BaBar's upper limit assuming $\widehat{\epsilon}_T=0$, left-hand side of figure~\ref{fig:Del}. Constraints on tensor interactions are weaker: $| \widehat{\epsilon}_T|\leq0.4$, assuming 
$\widehat{\epsilon}_S=0$ and BaBar's upper limit, right-hand side of figure~\ref{fig:Del}. Similar conclusions can be obtained 
for limits on the scalar coupling in the case of $\tau^- \to \pi^-\eta'\nu_{\tau}$ decays, see figures \ref{fig:Delp}. In this case $-0.011\leq \widehat{\epsilon}_S \leq 0.007$. 
It can be noticed that much looser limits are obtained for the tensor coupling in this case, $| \widehat{\epsilon}_T|\leq11$.

Secondly, constraints on scalar and tensor interactions can be set simultaneously from a comparison of experimental upper limits and eq. (\ref{percent}). This is represented in figures \ref{fig:DelC}, 
for the case of $\tau^- \to \pi^-\eta^{(\prime)}\nu_{\tau}$ decays. Clearly, the limits on the scalar and tensor couplings get slightly relaxed in this case with respect to the ones 
obtained when one of the couplings is assumed to vanish. These constraints can be largely improved at Belle II as it is shown in figures \ref{fig:DelPresc}, where we compare the limits 
that can be set on the $(\widehat{\epsilon}_S,\widehat{\epsilon}_T)$ plane by assuming that the branching ratio of $ \tau^- \to \pi^-\eta^{(\prime)}\nu_{\tau}$ can be measured with 50\% and 
20\% accuracy. Left (right)-hand side of figures \ref{fig:DelPresc} shows the sensitivity on the scalar and tensor couplings that can be obtained from improved measurements of the 
$\tau^-\to \pi^-\eta\nu_{\tau}( \tau^-\to \pi^-\eta'\nu_{\tau})$ branching fraction.

Table \ref{table1} summarizes the constraints on the scalar and tensor couplings that can be derived from the current upper limits on the branching ratios of 
$\tau^- \to \pi^-\eta^{(\prime)} \nu_{\tau}$ decays. We also display the constraints that can be obtained from forthcoming measurements of the branching fraction of these decays at Belle II 
experiment, by assuming a 20\% accuracy~\footnote{S. Descotes-Genon and B. Moussallam \cite{Orsay} pointed out that with this precision both in the measurement of the branching fraction of
$\tau^- \to \pi^-\eta \nu_{\tau}$ decays and in the theoretical knowledge of the participating scalar form factor, these decays will fix bounds on charged Higgs exchange competitive to those 
obtained from $B^-\to\tau^-\nu_\tau$ data.}.

\begin{table}
\begin{tabular}{|c|c|c|c|c|}
\hline
$\Delta$ & $\widehat{\epsilon}_S (\widehat{\epsilon}_T=0)$ & $\widehat{\epsilon}_T (\widehat{\epsilon}_S=0)$ & $\widehat{\epsilon}_S$ & $\widehat{\epsilon}_T$ \\
\hline \hline
\textbf{$\pi\eta$} &&&& \\ 
Babar  & $[-8.3,3.9]\cdot10^{-3}$& [-0.43,0.39] & $[-0.83,0.37]\cdot10^{-2}$ & [-0.55,0.50] \\
Belle  &$[-7.7,2.9]\cdot10^{-3}$&[-0.51,0.47]&$[-0.75,0.29]\cdot10^{-2}$&[-0.48,0.43] \\
CLEO  &$[-9.5,5.0]\cdot10^{-3}$&[-0.62,0.57]&$[-0.95,0.49]\cdot10^{-2}$&[-0.66,0.60] \\
Belle II &$([-4.8,2.0]\cdot10^{-3}$ &[-0.12,0.08]& $[-4.9,-4.3]\cdot10^{-3}\bigcup $& $[-0.20,-0.25]\bigcup $ \\
& & &$[-2.6,3.0]\cdot10^{-4}$ & [0.15,0.20] \\
\hline\hline 
\textbf{$\pi\eta^{\prime}$} &&&& \\ 
Babar  &$[-1.13,0.68]\cdot10^{-2}$ &$|\widehat{\epsilon}_T|<$11.4& $[-1.13,0.67]\cdot10^{-2}$&[-11.9,11.9] \\
Belle  & $[-1.07,0.60]\cdot10^{-2}$&$|\widehat{\epsilon}_T|<$10.6&$[-1.06,0.61]\cdot10^{-2}$& [-11.0,11.0] \\
Belle II &$[-4.8,2.3]\cdot10^{-3}$&[-1.35,1.41]&$[-4.8,-4.3]\cdot10^{-3}\bigcup$&$[-3.4,-2.7]\bigcup $\\
& & &$[-2.4,2.4]\cdot10^{-4}$ & $[2.7,3.3]$\\
\hline
\end{tabular}
\caption{\label{table1}Constraints on the scalar and tensor couplings obtained from current upper limits on the branching fractions and hypothetical measurements with 20\% accuracy at Belle II experiment.}
\end{table}

At this point it is interesting to compare the limits in Table \ref{table1} to those obtained in Ref.~\cite{Bhattacharya:2011qm} (see also \cite{Cirigliano:2012ab, Cirigliano:2013xha, Gonzalez-Alonso:2013ura}). 
For this we need to 
assume lepton universality because our study involves the $\tau$ flavor, while theirs electron and muon flavors. However, given the smallness of possible lepton universality 
violations, this is enough for current precision. It is clear that $\tau^- \to \pi^-\eta^{(\prime)}\nu_{\tau}$ decays are not competitive restricting tensor interactions. Our upper 
limits (using present data) are at the level of $|\widehat{\epsilon_T}|\lesssim 0.5$ while the radiative pion decay reaches the $10^{-4}$ level through Dalitz plot analysis \cite{Mateu:2007tr, Beldjoudi:1994hi, Cirigliano:2012ab, Cirigliano:2013xha, Gonzalez-Alonso:2013ura,Bychkov:2008ws}. 
On the contrary, our bounds are very 
competitive in the case of scalar interactions, where we get (with current data) $-0.009<|\widehat{\epsilon_S}|<0.004$, while $0^+\to0^+$ nuclear $\beta$ decays set limits (from the Fierz interference term) \cite{Hardy:2008gy}
at a few times $10^{-3}$~\footnote{As emphasized in e. g. Ref.~\cite{Cirigliano:2012ab}, if the flavor structure of the dynamics generating the non-standard interaction is known, then $R_\pi\equiv \Gamma(\pi\to e\nu (\gamma))/
\Gamma(\pi\to \mu\nu (\gamma))$ could provide the strongest constraint on $\widehat{\epsilon}_{S,T}$ (see also Refs.~\cite{Voloshin:1992sn, Herczeg:1994ur, Campbell:2003ir, Cirigliano:2007xi}).}. The potential of a precise measurement 
of these decays at Belle-II is illustrated in the very stringent bounds on $\widehat{\epsilon_S}$ appearing in table 1. For this, however,
it is crucial to improve our knowledge on the theoretical uncertainty of the scalar contribution~\footnote{Theoretical and experimental efforts in this direction can
be found in Refs.~\cite{Orsay,Escribano:2016ntp,Oller:1997ti,Furman:2002cg,Bugg:2008ig,Guo:2011pa,Guo:2012yt,Guo:2012ym,Adolph:2014rpp,Albaladejo:2015aca, 
Adolph:2015tqa,Dudek:2016cru,Guo:2016zep,Albaladejo:2017hhj}.}. Being quite conservative, we have re-calculated these constraints assuming that the 
scalar contribution to observables in the $\eta$ channel can be a factor seven smaller than quoted in \cite{Escribano:2016ntp} (like, for instance in Orsay's group prediction 
\cite{Orsay}) and this results in increasing the upper bound on $|\widehat{\epsilon_S}|$ one order of magnitude. Before results of Belle-II searches on these tau decays become available, 
more precise measurements of meson-meson scattering would be of enormous help in reducing the errors of the dominant scalar form factors, allowing thus the derivation of sharp limits on 
non-standard scalar interactions, as put forward in this article.
\section{Conclusions}

 The rare $\tau^- \to \pi^-\eta^{(\prime)}\nu_{\tau}$ decays, which are suppressed by G-parity in the Standard Model, can receive important contributions of New Physics. We have studied these decays 
 in the framework of the most general effective field theory which incorporate dimension-six operators and assumes left-handed neutrinos. We have found that the Dalitz plot, hadronic invariant mass 
 distribution and branching fraction are sensitive to the effects of scalar and tensor interactions and offer complementary information to the ones obtained from other low-energy processes.

 These decays will probably be observed for the first time at the Belle II experiment. The different observables studied in this paper will be very useful to characterize the underlying dynamics of 
 these decays. Our study indicates that these observables will be able to set very strong constraints on scalar interactions, or to set limits that are very competitive with other low-energy processes. 
 To the best of our knowledge, this is the first study aiming to disentangle SCC from G-parity violation in sensitive observables of tau lepton decays.

\acknowledgments {Work supported by CONACYT Project No. FOINS-296-2016 (`Fronteras de la Ciencia') and by projects 236394 and 250628  (`Ciencia B\'asica'). P.~R.~acknowledges discussions with Sergi Gonz\`alez-Sol\'is 
concerning numerical checks of the $\pi\eta^{(\prime)}$ scalar form factors. We thank very much useful discussions with Mart\'in Gonz\'alez-Alonso.}



\begin{thebibliography}{37}

\bibitem{Lee:1956sw} 
  T.~D.~Lee and C.~N.~Yang,
  Nuovo Cim.\  {\bf 10}, 749 (1956).

\bibitem{Leroy:1977pq} 
  C.~Leroy and J.~Pestieau,
  Phys.\ Lett.\  {\bf 72B}, 398 (1978).

\bibitem{Weinberg:1958ut} 
  S.~Weinberg,
  Phys.\ Rev.\  {\bf 112}, 1375 (1958).

\bibitem{Branco:2011iw}
  G.~C.~Branco, P.~M.~Ferreira, L.~Lavoura, M.~N.~Rebelo,
M.~Sher and J.~P.~Silva,
  Phys.\ Rept.\  {\bf 516}, 1 (2012)

  \bibitem{Jung:2010ik} 
  M.~Jung, A.~Pich and P.~Tuz\'on,
 JHEP {\bf 1011}, 003 (2010).


\bibitem{Becirevic:2016yqi} 
D.~Becirevic, S.~Fajfer, N.~Kosnik and O.~Sumensari,
Phys.\ Rev.\ D {\bf 94}, 115021 (2016).

  \bibitem{Severijns:2006dr}
  N.~Severijns, M.~Beck and O.~Naviliat-Cuncic,
  Rev.\ Mod.\ Phys.\  {\bf 78} (2006) 991.

\bibitem{Triambak:2017jpw}
  S.~Triambak {\it et al.},
  Phys.\ Rev.\ C {\bf 95} (2017) ,  035501
   Addendum: [Phys.\ Rev.\ C {\bf 95} (2017),  049901].

  \bibitem{BRs}
    Y.~Meurice,
  Phys.\ Rev.\ D {\bf 36}, 2780 (1987);
  A.~Bramon, S.~Narison and A.~Pich,
  Phys.\ Lett.\ B {\bf 196} (1987) 543;
  A.~Pich,
  Phys.\ Lett.\ B {\bf 196} (1987) 561;
  J.~L.~D\'iaz-Cruz and G.~L\'opez Castro,
  Mod.\ Phys.\ Lett.\ A {\bf 6}, 1605 (1991);
 S.~Nussinov and A.~Soffer,
  Phys.\ Rev.\ D {\bf 78}, 033006 (2008),
  Phys.\ Rev.\ D {\bf 80}, 033010 (2009);
  N.~Paver and Riazuddin,
  Phys.\ Rev.\ D {\bf 82}, 057301 (2010), 
  Phys.\ Rev.\ D {\bf 84}, 017302 (2011);
 M.~K.~Volkov and D.~G.~Kostunin,
  Phys.\ Rev.\ D {\bf 86}, 013005 (2012);
  
  \bibitem{Vienna}
 H.~Neufeld and H.~Rupertsberger,
  Z.\ Phys.\ C {\bf 68}, 91 (1995).
  
  \bibitem{Orsay}
  S.~Descotes-Genon and B.~Moussallam,
  Eur.\ Phys.\ J.\ C {\bf 74} (2014) 2946.
    
\bibitem{Escribano:2016ntp} 
  R.~Escribano, S.~Gonz\`alez-Sol\'is and P.~Roig,
  Phys.\ Rev.\ D {\bf 94},  034008 (2016).
  
  \bibitem{Guevara:2016trs}
  A.~Guevara, G.~L\'opez-Castro and P.~Roig,
  Phys.\ Rev.\ D {\bf 95} (2017),  054015.
  
  \bibitem{Hernandez-Tome:2017pdc}
  G.~Hern\'andez-Tom\'e, G.~L\'opez-Castro and P.~Roig,
  arXiv:1707.03037 [hep-ph]. To be published in Phys. Rev. D.
  
\bibitem{delAmoSanchez:2010pc} 
  P.~del Amo Sanchez {\it et al.} [BaBar Collaboration],
  Phys.\ Rev.\ D {\bf 83}, 032002 (2011).
  
\bibitem{Hayasaka:2009zz} 
  K.~Hayasaka [Belle Collaboration],
  PoS EPS {\bf -HEP2009}, 374 (2009).
  
  \bibitem{Bartelt:1996iv} 
  J.~E.~Bartelt {\it et al.} [CLEO Collaboration],
  Phys.\ Rev.\ Lett.\  {\bf 76}, 4119 (1996).

  \bibitem{Aubert:2008nj}
  B.~Aubert {\it et al.} [BaBar Collaboration],
  Phys.\ Rev.\ D {\bf 77} (2008) 112002.
 
  \bibitem{Bergfeld:1997zt}
  T.~Bergfeld {\it et al.} [CLEO Collaboration],
  Phys.\ Rev.\ Lett.\  {\bf 79} (1997) 2406.
  
\bibitem{Abe:2010gxa} 
  T.~Abe {\it et al.} [Belle-II Collaboration],
  arXiv:1011.0352 [physics.ins-det].

  \bibitem{B2TIPReport}
Belle-II Physics Book, Belle-II Collaboration and B2TIP-Community, to be published in Progress of Theoretical and Experimental Physics.
  
  \bibitem{Buchmuller:1985jz}
  W.~Buchmuller and D.~Wyler,
  Nucl.\ Phys.\ B {\bf 268} (1986) 621.

  \bibitem{Grzadkowski:2010es}
  B.~Grzadkowski, M.~Iskrzynski, M.~Misiak and J.~Rosiek,
  JHEP {\bf 1010} (2010) 085.
  
\bibitem{Bhattacharya:2011qm} 
  T.~Bhattacharya, V.~Cirigliano, S.~D.~Cohen, A.~Filipuzzi, M.~Gonz\'alez-Alonso, M.~L.~Graesser, R.~Gupta and H.~W.~Lin,
  Phys.\ Rev.\ D {\bf 85}, 054512 (2012).

\bibitem{Cirigliano:2009wk} 
  V.~Cirigliano, J.~Jenkins and M.~Gonz\'alez-Alonso,
  Nucl.\ Phys.\ B {\bf 830}, 95 (2010).

   \bibitem{Chang:2014iba}
  H.~M.~Chang, M.~Gonz\'alez-Alonso and J.~Mart\'in Camalich,
  Phys.\ Rev.\ Lett.\  {\bf 114} (2015) no.16,  161802.

  \bibitem{Gonzalez-Alonso:2016etj}
  M.~Gonz\'alez-Alonso and J.~Mart\'in Camalich,
  JHEP {\bf 1612} (2016) 052.

  \bibitem{Gonzalez-Alonso:2016sip}
  M.~Gonz\'alez-Alonso and J.~Mart\'in Camalich,
  arXiv:1606.06037 [hep-ph].
  
   \bibitem{Erler:2002mv} 
A. Sirlin, Rev. Mod. Phys. {\bf 50}, 573 (1978); Nucl. Phys. {\bf B71}, 29 (1974);
 W. J. Marciano and A. Sirlin, Phys. Rev. Lett. {\bf 61}, 1815 (1988),  {\it ibid.} {\bf 71}, 3629 (1993);
W. J. Marciano and A. Sirlin, Phys. Rev. Lett. {\bf 56}, 22 (1986); 
A. Sirlin, Nucl. Phys. {\bf B196}, 83 (1982); 
  J.~Erler,  Rev.\ Mex.\ Fis.\  {\bf 50}, 200 (2004).
  
  \bibitem{ChPT}
  A.~Pich,
  Rept.\ Prog.\ Phys.\  {\bf 58} (1995) 563;
  G.~Ecker,
  Prog.\ Part.\ Nucl.\ Phys.\  {\bf 35} (1995) 1.

  \bibitem{Aoki:2016frl}
  S.~Aoki {\it et al.},
  Eur.\ Phys.\ J.\ C {\bf 77} (2017) no.2,  112.

  
  \bibitem{Weinberg:1978kz}
  S.~Weinberg,
  Physica A {\bf 96} (1979) 327.

  \bibitem{Gasser:1983yg}
  J.~Gasser and H.~Leutwyler,
  Annals Phys.\  {\bf 158} (1984) 142.
 
  \bibitem{Gasser:1984gg}
  J.~Gasser and H.~Leutwyler,
  Nucl.\ Phys.\ B {\bf 250} (1985) 465.

\bibitem{Colangelo:1999kr} 
  G.~Colangelo, G.~Isidori and J.~Portol\'es,
  Phys.\ Lett.\ B {\bf 470}, 134 (1999).
  
  \bibitem{PDG}
  {C.~Patrignani {\it et al.} [Particle Data Group],
  Chin.\ Phys.\ C {\bf 40} (2016) no.10,  100001.
  }
  
  
  \bibitem{Cata:2008zc}
 {O.~Cat\`a and V.~Mateu,
  Phys.\ Rev.\ D {\bf 77} (2008) 116009.
  }
  
  \bibitem{Becirevic:2003pn}
 {D.~Becirevic, V.~Lubicz, F.~Mescia and C.~Tarantino,
  JHEP {\bf 0305} (2003) 007.
 }
 
 \bibitem{Braun:2003jg}
 {V.~M.~Braun, T.~Burch, C.~Gattringer, M.~Gockeler, G.~Lacagnina, S.~Schaefer and A.~Schafer,
  Phys.\ Rev.\ D {\bf 68} (2003) 054501.
 }
 
 \bibitem{Donnellan:2007xr}
  { M.~A.~Donnellan {\it et al.},
  PoS LAT {\bf 2007} (2007) 369.
  }
  
\bibitem{Cata:2007ns} 
  O.~Cat\`a and V.~Mateu,
  JHEP {\bf 0709}, 078 (2007).
  
    \bibitem{ChPTLargeN}
  H.~Leutwyler,
  Nucl.\ Phys.\ Proc.\ Suppl.\  {\bf 64} (1998) 223;
  R.~Kaiser and H.~Leutwyler,
  In *Adelaide 1998, Nonperturbative methods in quantum field theory* 15-29;
 R.~Kaiser and H.~Leutwyler,
  Eur.\ Phys.\ J.\ C {\bf 17} (2000) 623.

  \bibitem{2anglemixing}
  J.~Schechter, A.~Subbaraman and H.~Weigel,
  Phys.\ Rev.\ D {\bf 48} (1993) 339;
  T.~Feldmann, P.~Kroll and B.~Stech,
  Phys.\ Rev.\ D {\bf 58} (1998) 114006;
  Phys.\ Lett.\ B {\bf 449} (1999) 339; 
  T.~Feldmann,
  Int.\ J.\ Mod.\ Phys.\ A {\bf 15} (2000) 159.

\bibitem{Mateu:2007tr} 
  V.~Mateu and J.~Portol\'es,
  Eur.\ Phys.\ J.\ C {\bf 52}, 325 (2007).
  O.~Cata and V.~Mateu,
  Phys.\ Rev.\ D {\bf 77}, (2008) 116009.


  \bibitem{Beldjoudi:1994hi}
  L.~Beldjoudi and T.~N.~Truong,
  Phys.\ Lett.\ B {\bf 351} (1995) 357.
 
  \bibitem{Cirigliano:2012ab}
  V.~Cirigliano, M.~Gonz\'alez-Alonso and M.~L.~Graesser,
  JHEP {\bf 1302} (2013) 046.
  
  \bibitem{Cirigliano:2013xha}
 {
  V.~Cirigliano, S.~Gardner and B.~Holstein,
  Prog.\ Part.\ Nucl.\ Phys.\  {\bf 71} (2013) 93.
}

  \bibitem{Gonzalez-Alonso:2013ura}
  M.~Gonz\'alez-Alonso and J.~Mart\'in Camalich,
  Phys.\ Rev.\ Lett.\  {\bf 112} (2014),  042501.
  
  \bibitem{Bychkov:2008ws}
  M.~Bychkov {\it et al.},
  Phys.\ Rev.\ Lett.\  {\bf 103} (2009) 051802.

  \bibitem{Hardy:2008gy}
  J.~C.~Hardy and I.~S.~Towner,
  Phys.\ Rev.\ C {\bf 79} (2009) 055502.
  
  \bibitem{Voloshin:1992sn}
  M.~B.~Voloshin,
  Phys.\ Lett.\ B {\bf 283} (1992) 120.
  
  \bibitem{Herczeg:1994ur}
  P.~Herczeg,
  Phys.\ Rev.\ D {\bf 49} (1994) 247.
  
  \bibitem{Campbell:2003ir}
  B.~A.~Campbell and D.~W.~Maybury,
  Nucl.\ Phys.\ B {\bf 709} (2005) 419.
  
  \bibitem{Cirigliano:2007xi}
  V.~Cirigliano and I.~Rosell,
  Phys.\ Rev.\ Lett.\  {\bf 99} (2007) 231801
  
  \bibitem{Oller:1997ti}
   {J.~A.~Oller and E.~Oset,
  Nucl.\ Phys.\ A {\bf 620} (1997) 438
   Erratum: [Nucl.\ Phys.\ A {\bf 652} (1999) 407].
}
  \bibitem{Furman:2002cg}
   {A.~Furman and L.~Lesniak,
  Phys.\ Lett.\ B {\bf 538} (2002) 266.
  }
  \bibitem{Bugg:2008ig}
   {D.~V.~Bugg,
  Phys.\ Rev.\ D {\bf 78} (2008) 074023.
  }
  \bibitem{Guo:2011pa}
   {Z.~H.~Guo and J.~A.~Oller,
  Phys.\ Rev.\ D {\bf 84} (2011) 034005.
  }
  \bibitem{Guo:2012yt}
   {Z.~H.~Guo, J.~A.~Oller and J.~Ruiz de Elvira,
  Phys.\ Rev.\ D {\bf 86} (2012) 054006.
  }
  \bibitem{Guo:2012ym}
   {Z.~H.~Guo, J.~A.~Oller and J.~Ruiz de Elvira,
  Phys.\ Lett.\ B {\bf 712} (2012) 407.
  }
  \bibitem{Adolph:2014rpp}
   {C.~Adolph {\it et al.} [COMPASS Collaboration],
  Phys.\ Lett.\ B {\bf 740} (2015) 303.
  }
  \bibitem{Albaladejo:2015aca}
   {M.~Albaladejo and B.~Moussallam,
  Eur.\ Phys.\ J.\ C {\bf 75} (2015) no.10,  488.
  }
  \bibitem{Adolph:2015tqa}
   {C.~Adolph {\it et al.} [COMPASS Collaboration],
  Phys.\ Rev.\ D {\bf 95} (2017) no.3,  032004.
  }
  \bibitem{Dudek:2016cru}
   {J.~J.~Dudek {\it et al.} [Hadron Spectrum Collaboration],
  Phys.\ Rev.\ D {\bf 93} (2016) no.9,  094506.
  }
  \bibitem{Guo:2016zep}
   {Z.~H.~Guo, L.~Liu, U.~G.~Meißner, J.~A.~Oller and A.~Rusetsky,
  Phys.\ Rev.\ D {\bf 95} (2017) no.5,  054004.
  }
  \bibitem{Albaladejo:2017hhj}
   {M.~Albaladejo and B.~Moussallam,
  Eur.\ Phys.\ J.\ C {\bf 77} (2017) no.8,  508.
 } 
  \end{thebibliography}
\end{document}